\documentclass[a4paper,11pt]{article}

\usepackage{amsmath,amsthm,amsbsy,amssymb,amsfonts,graphicx,mathrsfs,bm,cite,color}
\usepackage{hyperref}

\makeatletter
\catcode`\@=11
\@addtoreset{equation}{section}

\makeatother

\makeatletter

\@addtoreset{table}{section}
\makeatother

\renewcommand{\thefootnote}{\arabic{footnote}}

\newcommand{\Exp}[1]{\operatorname{e}^{#1}}

\newcommand{\abs}[1]{\lvert {#1} \rvert}
\newcommand{\rmd}{{\mathrm{d}}}
\newcommand{\nn}{\nonumber}
\newcommand{\Lie}{\pounds}
\newcommand{\gLie}{\hat{\pounds}}

\newcommand{\cA}{\mathcal A}\newcommand{\cB}{\mathcal B}
\newcommand{\cC}{\mathcal C}\newcommand{\cD}{\mathcal D}
\newcommand{\cE}{\mathcal E}\newcommand{\cF}{\mathcal F}
\newcommand{\cH}{\mathcal H}
\newcommand{\cI}{\mathcal I}\newcommand{\cJ}{\mathcal J}
\newcommand{\cK}{\mathcal K}\newcommand{\cL}{\mathcal L}
\newcommand{\cM}{\mathcal M}

\newcommand{\lR}{\mathbb{R}}

\newcommand{\sfa}{\mathsf{a}}
\newcommand{\sfb}{\mathsf{b}}
\newcommand{\sfc}{\mathsf{c}}
\newcommand{\sff}{\mathsf{f}}
\newcommand{\sfA}{\mathsf{A}}
\newcommand{\sfK}{\mathsf{K}}
\newcommand{\sfR}{\mathsf{R}}
\newcommand{\sfT}{\mathsf{T}}

\newcommand{\SL}{\text{SL}}
\newcommand{\SO}{\text{SO}}
\newcommand{\GL}{\text{GL}}
\newcommand{\OO}{\text{O}}

\allowdisplaybreaks[3]

\begin{document}

\begin{titlepage}
\renewcommand{\thefootnote}{\fnsymbol{footnote}}

\vspace*{1cm}

\centerline{\Large\textbf{$U$-duality extension of Drinfel'd double}}%

\vspace{1.5cm}

\centerline{\large Yuho Sakatani}

\vspace{0.2cm}

\begin{center}
{\it Department of Physics, Kyoto Prefectural University of Medicine,}\\
{\it Kyoto 606-0823, Japan}\\
{\small\texttt{yuho@koto.kpu-m.ac.jp}}
\end{center}

\vspace*{2mm}

\begin{abstract}
A family of algebras $\mathcal{E}_n$ that extends the Lie algebra of the Drinfel'd double is proposed. This allows us to systematically construct the generalized frame fields $E_A{}^I$ which realize the proposed algebra by means of the generalized Lie derivative, i.e., $\hat{\pounds}_{E_A}E_B{}^I = - \mathcal{F}_{AB}{}^C\,E_C{}^I$. By construction, the generalized frame fields include a twist by a Nambu--Poisson tensor. A possible application to the non-Abelian extension of $U$-duality and a generalization of the Yang--Baxter deformation are also discussed. 
\end{abstract}

\thispagestyle{empty}
\end{titlepage}

\setcounter{footnote}{0}

\newpage

\section{Introduction}

The familiar $T$-duality is a symmetry of string theory when the target space has commuting (or Abelian) Killing vectors. 
An extension of the $T$-duality, where the Killing vectors do not commute with each other, was proposed in \cite{Fridling:1983ha,Fradkin:1984ai,hep-th:9210021,hep-th:9308154,hep-th:9308112,hep-th:9403155}, and it is known as non-Abelian $T$-duality (see \cite{1903.12175} for a list of references). 
Subsequently, a further extension, called the Poisson--Lie (PL) $T$-duality was found in \cite{hep-th:9502122,hep-th:9509095}, and it can be applied to a more general class of target spaces. 
As has been discovered there, there is a group structure of the Drinfel'd double behind the PL $T$-duality, and the symmetry of the PL $T$-duality can be understood as a freedom in the choice of the physical subgroup $G$ out of the doubled Lie group. 

In the case of Abelian $T$-duality, the double extension of the target space has been proposed in various contexts (see for example \cite{Duff:1989tf,Tseytlin:1990nb,Tseytlin:1990va,Tseytlin:1990hn,hep-th:9201040}). 
The geometry of the doubled space was studied in \cite{hep-th:9302036,hep-th:9305073,hep-th:9308133}, and more recently, the idea has been developed in the context of double field theory (DFT) \cite{0904.4664,1006.4823}. 
In the original formulation of DFT, the symmetry of Abelian $T$-duality is manifest, but the non-Abelian $T$-duality or the PL $T$-duality has not been clearly discussed. 
A new formulation of DFT on group manifolds (called DFT$_{\text{WZW}}$) has been developed in \cite{1410.6374,1502.02428,1509.04176}, and in the recent works \cite{1707.08624,1810.11446}, the PL $T$-duality has been studied in the framework of DFT$_{\text{WZW}}$. 
In more recent papers \cite{1903.12175,1904.00362}, the non-Abelian $T$-duality and the PL $T$-duality have been discussed by using another approach, called the gauged DFT \cite{1109.0290,1109.4280,1201.2924,1304.1472,1705.08181,1706.08883} (see also \cite{1901.04777,1910.09997} for recent discussion on the Drinfel'd double and related aspects in DFT). 
Thus, DFT is now not restricted to Abelian $T$-duality but can also be applied to the non-Abelian extensions. 

When the target space has $D$ Abelian Killing vectors, the $T$-duality group of type II superstring theory is $\OO(D,D)$. 
As is well-known, this $T$-duality group is only a subgroup of a larger duality group, called the $U$-duality group. 
The $U$-duality group is $E_n$ ($n\equiv D+1$), which is summarized in Table \ref{table:En}. 
\begin{table}[b]
\centerline{\begin{tabular}{|c||c|c|c|c|c|c|c|}\hline
 $n$ & 2 & 3 & 4 & 5 & 6 & 7 & 8 \\\hline\hline
 $E_n$ & $\SL(2)\times \lR^+$ & $\SL(3)\times \SL(2)$ & $\SL(5)$ & $\SO(5,5)$ & $E_6$ & $E_7$ & $E_8$ \\\hline
 $D_n$ & 3 & 6 & 10 & 16 & 27 & 56 & 248 \\\hline
 $d_n$ & 2 & 3 & 5 & 10 & 27 & 133 & 3875$+$1 \\\hline
\end{tabular}}
\caption{The $U$-duality group $E_n$ for each $n$ and dimensions of two representations, known as the $R_1$-representation and the $R_2$-representation.
\label{table:En}}
\end{table}
In order to manifest the $U$-duality symmetry, the doubled space is not enough. 
As it has been discussed in \cite{Duff:1990hn,hep-th:0307098,0712.1795,0902.1509,1006.0893,1008.1763}, we need to extend the $n$-dimensional space (with Abelian Killing vectors) into an extended space with dimension $D_n$, which is the dimension of the vector representation $R_1$ of the $U$-duality group (see Table \ref{table:En}). 
For higher $n$, it is much larger than the doubled space with dimension $2D$, and the extended space is called the exceptional space for the obvious reason. 
The $U$-duality-manifest formulation of supergravities have been studied in \cite{hep-th:0104081,hep-th:0402140,hep-th:0511153,0705.0752,0712.1795,1008.1763,1103.5733,1111.0459,1111.1642,1112.3989,1206.7045,1212.1586}, and more recently, a formulation using a similar language to DFT has been developed in \cite{1110.3930,1208.5884,1302.5419,1308.1673,1312.0614,1312.4542,1312.4549,1406.3348}, which is called the exceptional field theory (EFT). 
Similar to the case of DFT, for the consistency of the theory, we need to choose a physical subspace from the extended space, and all of the supergravity fields are defined on the physical subspace. 
In the case of DFT, the maximal dimension allowed for the consistency is always $D$-dimensional, but in the case of EFT, there are two maximal choices, $n$-dimensions and $D(=n-1)$-dimensions \cite{1311.5109}. 
If we adopt the former choice, the target space of M-theory is reproduced while the latter reproduces that of type IIB theory \cite{1311.5109}. 
In this sense, EFT unifies the geometry of M-theory and type IIB string theory, and the structure of the exceptional space is much richer than that of the double space. 

Unlike the case of $T$-duality, there is no concrete proposal for the extension of $U$-duality when the Killing vectors are non-Abelian. 
Originally, the non-Abelian $T$-duality was discovered by introducing certain gauge fields (associated with the Killing vectors) into the string sigma model. 
This allows us to reformulate the string theory as a gauged sigma model, which reduces to the standard string sigma model if we first eliminate certain auxiliary fields. 
On the other hand, if we eliminate the gauge fields first, the string sigma model on the dual geometry is recovered \cite{hep-th:9210021,hep-th:9308154,hep-th:9308112,hep-th:9403155}, and, in this sense, the gauged sigma model connects the original geometry and the dual geometry. 
If we try to apply the same procedure to the membrane sigma model, we face a difficulty (see \cite{1903.12175}). 
In order to formulate the non-Abelian extension of $U$-duality, the approach of the PL $T$-duality will be more useful. 
For this purpose, we need to extend the exceptional space to some extended group manifold, similar to the Drinfel'd double. 
Such an extension has been studied in \cite{1705.09304}, but the relation to the Drinfel'd double is not so clear and no proposal has been made for the extension of the PL $T$-duality. 

In this paper, by using the idea of EFT, we propose a family of Leibniz algebra $\cE_n$ which contains the Lie algebra of the Drinfel'd double as a subalgebra in a particular case. 
For simplicity, we restrict our analysis to the case $n\leq 4$\,. 
In that case, the generators of the algebra $T_A$ ($A=1,\dotsc,D_n$) can be parameterized as $\{T_A\}=\{T_a,\,T^{a_1a_2}\}$, where $a=1,\dotsc,n$ and $T^{a_1a_2}=-T^{a_2a_1}$. 
The algebra can be expressed as
\begin{align}
\begin{split}
 T_a\circ T_b &= f_{ab}{}^c\,T_c \,,
\\
 T_a\circ T^{b_1b_2} &= f_a{}^{b_1b_2c}\,T_c + 2\,f_{ac}{}^{[b_1}\, T^{b_2]c} \,,
\\
 T^{a_1a_2}\circ T_b &= -f_b{}^{a_1a_2 c}\,T_c + 3\,f_{[c_1c_2}{}^{[a_1}\,\delta^{a_2]}_{b]} \,T^{c_1c_2}\,,
\\
 T^{a_1a_2}\circ T^{b_1b_2} &= -2\, f_d{}^{a_1a_2[b_1}\, T^{b_2]d}\,,
\end{split}
\label{eq:En-algebra}
\end{align}
where $f_{ab}{}^c=f_{[ab]}{}^c$ and $f_a{}^{b_1b_2b_3}=f_a{}^{[b_1b_2b_3]}$\,, and the following bilinear forms are defined:
\begin{align}
 \langle T_a,\, T^{b_1b_2}\rangle_c = 2!\,\delta^{[b_1}_{a}\,\delta^{b_2]}_c \,,\qquad 
 \langle T^{a_1a_2},\, T^{b_1b_2}\rangle_{c_1\cdots c_4} = 4!\,\delta^{a_1}_{[c_1}\,\delta^{a_2}_{c_2}\,\delta^{b_1}_{c_3}\,\delta^{b_2}_{c_4]}\,,
\label{eq:section}
\end{align}
which naturally extend the standard bilinear form $\langle \cdot,\,\cdot \rangle$ of the Drinfel'd double. 

By using the proposed algebra, we can systematically construct the generalized frame fields $E_A{}^I(x)$ which satisfy
\begin{align}
 \gLie_{E_A} E_B{}^I = - \cF_{AB}{}^C\,E_C{}^I\,,
\label{eq:gen-Lie-E}
\end{align}
where $\cF_{AB}{}^C$ is the structure constant of the proposed Leibniz algebra $T_A\circ T_B = \cF_{AB}{}^C\,T_C$\,.
Here, $\gLie_{V}$ is the generalized Lie derivative in EFT, which generates the gauge transformations in EFT. 
The systematic construction of $E_A{}^I$ is indispensable to the performance of the PL $T$-duality or its extension PL $T$-plurality \cite{hep-th:0205245}, and we expect that the $U$-duality extension presented in this paper will be useful in studying the non-Abelian extension of $U$-duality. 

The structure of this paper is as follows. 
In section \ref{sec:PL-T}, we briefly summarize the idea of the PL $T$-duality. 
In particular, we explain how the relation \eqref{eq:gen-Lie-E} is important in the PL $T$-duality. 
In section \ref{sec:algebra}, we find the algebra $\cE_n$ and study its detailed properties. 
The construction of the generalized frame fields is explained in section \ref{sec:frame}. 
In section \ref{sec:examples}, we show several examples of the $\cE_n$ algebra. 
Section \ref{sec:discussion} is devoted to the summary and discussion. 

\section{Poisson--Lie $T$-duality}
\label{sec:PL-T}

In this section, we review the PL $T$-duality by using the language of DFT. 

\paragraph{Basics definitions in DFT:}
Let us set up basic definitions of DFT. 
We consider a doubled space which has the generalized coordinates $(x^M)=(x^m,\,\tilde{x}_m)$ ($m=1,\dotsc,D$). 
The metric and the $B$-field are packaged into the generalized metric,
\begin{align}
 (\cH_{MN}) = \begin{pmatrix} g_{mn}-B_{mp}\,g^{pq}\,B_{qn} & -B_{mp}\,g^{pn} \\ g^{mp}\,B_{pn} & g^{mn} \end{pmatrix},
\label{eq:cH-DFT}
\end{align}
and the dilaton $\Phi$ is redefined into the $T$-duality-invariant combination,
\begin{align}
 \Exp{-2d} \equiv \Exp{-2\Phi}\sqrt{\abs{g}}\,,
\end{align}
where $d(x)$ is called the DFT dilaton. 
We denote the $\OO(D,D)$-invariant metric as
\begin{align}
 (\eta_{MN}) \equiv \begin{pmatrix} 0 & \delta_m^n \\ \delta^m_n & 0 \end{pmatrix}, \qquad 
 (\eta^{MN}) \equiv \begin{pmatrix} 0 & \delta^m_n \\ \delta_m^n & 0 \end{pmatrix},
\end{align}
and use these to raise or lower the indices $M,N$. 
The fields $\cH_{MN}(x)$ and $d(x)$ are formally defined on the doubled space, but for consistency we impose the section condition
\begin{align}
 \eta^{MN}\,\partial_M A(x)\,\partial_N B(x) = 0 \,, \qquad
 \eta^{MN}\,\partial_M \partial_N A(x) = 0 \,, 
\label{eq:SC-DFT}
\end{align}
for arbitrary fields $A(x)$ and $B(x)$\,. 
According to the section condition, all of the fields can depend only on a set of $D$ coordinates, and in this paper we choose $x^m$ as such $D$ coordinates. 
Any other choices can be mapped to this choice by performing a $T$-duality transformation. 

A generalization of the Lie derivative, called the generalized Lie derivative, is defined as
\begin{align}
 \gLie_V W^M \equiv V^N\,\partial_N W^M - \bigl(\partial_N V^M - \partial^M V_N\bigr)\, W^N\,,
\end{align}
which generates the gauge transformations in DFT. 
Under the section where all fields depend only on the physical coordinates $x^m$, the generalized Lie derivative reduces to
\begin{align}
 \bigl(\gLie_V W^M\bigr) = \begin{pmatrix} \Lie_v w^m \\ (\Lie_v \tilde{w}_1 - \iota_w \rmd \tilde{v}_1)_m \end{pmatrix},
\end{align}
where we have parameterized the generalized vectors as $(V^M)=(v^m,\,\tilde{v}_m)$ and $(W^M)=(w^m,\,\tilde{w}_m)$ and denoted the 1-forms as $\tilde{v}_1\equiv \tilde{v}_m\,\rmd x^m$ and $\tilde{w}_1\equiv \tilde{w}_m\,\rmd x^m$. 

\paragraph{Abelian $T$-duality:}
Now, let us consider the $T$-duality by using the above notation. 
In order to perform the standard Abelian $T$-duality, the generalized metric and the DFT dilaton are required to be constant,
\begin{align}
 \cH_{MN}(x)=\hat{\cH}_{MN}\,,\qquad d(x)=\hat{d} \,,
\label{eq:Abelian-dualizable-BG}
\end{align}
in a certain adapted coordinate system. 
In this constant background, equations of motion of DFT are trivially satisfied, and a constant $\OO(D,D)$ transformation
\begin{align}
 \hat{\cH}'_{MN} = C_M{}^P\,C_N{}^Q\,\hat{\cH}_{PQ}\,,\qquad 
 \hat{d}'= \hat{d} 
\label{eq:Ableian-T-dual}
\end{align}
maps the solution to another constant solution. 
This is the Abelian $T$-duality. 

\paragraph{Poisson--Lie $T$-dualizable backgrounds:}
When we consider the PL $T$-duality, the target space is allowed to be non-constant. 
The generalized metric $\cH_{MN}$ can be twisted by a non-constant matrix $E_M{}^A(x)$ and the DFT dilaton also can have a non-constant factor,
\begin{align}
 \cH_{MN}(x) = E_M{}^A(x)\,E_N{}^B(x)\, \hat{\cH}_{AB} \,,\qquad \Exp{-2d(x)} = \abs{\ell(x)} \Exp{-2\hat{d}}\,,
\label{eq:PL-dualizable-BG}
\end{align}
where $\abs{\ell(x)}\equiv \abs{\det(\ell_m^a)}$ and the matrices, $E_M{}^A(x)$ and $\ell_m^a(x)$ are defined as follows. 
First, we introduce the Lie algebra of the Drinfel'd double,
\begin{align}
 [T_a,\,T_b] = f_{ab}{}^c\,T_c\,,\qquad
 [T_a,\,\tilde{T}^b] = \tilde{f}^{bc}{}_a\,T_c - f_{ac}{}^b\,\tilde{T}^c\,,\qquad
 [\tilde{T}^a,\,\tilde{T}^b] = \tilde{f}^{ab}{}_c\,\tilde{T}^c\,,
\label{eq:Drinfeld-algebra}
\end{align}
which is equipped with the $ad$-invariant bilinear form,
\begin{align}
 \langle T_A,\,T_B \rangle = \eta_{AB}\,,\qquad 
 (\eta_{AB}) = \begin{pmatrix} 0 & \delta_a^b \\ \delta^a_b & 0 \end{pmatrix}, \qquad 
 (T_A) \equiv (T_a,\,\tilde{T}^a)\,.
\end{align}
The indices $A,B$ are raised or lowered by using the metric $\eta_{AB}$\,. 
The structure constants $f_{ab}{}^c$ and $\tilde{f}^{ab}{}_c$ can be chosen arbitrarily as long as Jacobi identities are satisfied. 
Secondly, we decompose the algebra into two subalgebras $\mathfrak{g}\oplus\tilde{\mathfrak{g}}$, where $\mathfrak{g}$ is the ``physical algebra'' spanned by $T_a$ and the dual algebra $\tilde{\mathfrak{g}}$ is spanned by $\tilde{T}^a$\,. 
Each of these is maximally isotropic for the bilinear form $\langle \cdot , \cdot \rangle$. 
We then define a group element $g$ by using the generators of the physical subalgebra, for example, $g=\Exp{x^a\,T_a}$\,, and define the left- and right-invariant 1-forms as
\begin{align}
 \ell \equiv \ell^a_m\,T_a\,\rmd x^m \equiv g^{-1}\,\rmd g\,,\qquad 
 r \equiv r^a_m\,T_a\,\rmd x^m \equiv \rmd g\,g^{-1} \,.
\end{align}
The left- and right-invariant vectors are denoted by $v_a^m$ and $e_a^m$ ($\ell^a_m\,v^m_b = \delta^a_b$ and $r^a_m\,e^m_b = \delta^a_b$). 
We also parameterize the adjoint action of $g^{-1}$ on the generators of the Drinfel'd double as
\begin{align}
 g^{-1}\,T_A\,g \equiv M_A{}^B\,T_B \,,\qquad M \equiv \begin{pmatrix} \delta_a^c & 0 \\ -\Pi^{ac} & \delta^a_c \end{pmatrix} \begin{pmatrix} a_c{}^b & 0 \\ 0 & (a^{-1})_b{}^c \end{pmatrix} ,
\end{align}
where we have used the fact that the structure constants \eqref{eq:Drinfeld-algebra} show that $M$ is an $\OO(D,D)$ matrix of the lower-triangular form, and thus $\Pi^{ab}$ is an anti-symmetric matrix while $a_c{}^b$ is an element of $\GL(D)$. 
Finally, by using the above quantities, we define the twist matrix as
\begin{align}
 (E_M{}^A) \equiv \begin{pmatrix} r_m^a & 0 \\ -e^m_c\,\Pi^{c a} & e^m_a \end{pmatrix}. 
\label{eq:twist-E}
\end{align}
Once the structure constants $f_{ab}{}^c$ and $\tilde{f}^{ab}{}_c$ and the parameterization of $g$ are given, the matrices $E_M{}^A(x)$ and $\ell_m^a(x)$ are uniquely obtained. 
Then, by using these matrices, the PL $T$-dualizable background is expressed as \eqref{eq:PL-dualizable-BG}. 
For later convenience, it is useful to note that at the identity $g=1$ (which corresponds to $x^a=0$ when $g=\Exp{x^a\,T_a}$), we have
\begin{align}
 \Pi^{ab}(x)\big\rvert_{g=1} = 0\,,\qquad a_a{}^b(x)\big\rvert_{g=1}=\delta_a^b \,,
\label{eq:Pi=0}
\end{align}
by their definitions. 
We also note that the Abelian $T$-dualizable background, i.e., the constant background \eqref{eq:Abelian-dualizable-BG}, is reproduced as a particular case, $f_{ab}{}^c=0$ and $\tilde{f}^{ab}{}_c=0$ with $g=\Exp{x^a\,T_a}$\,. 

\paragraph{Poisson--Lie $T$-duality:}
We denote the Lie algebra of the Drinfel'd double as
\begin{align}
 [T_A,\,T_B]=\cF_{AB}{}^C\,T_C\,,
\end{align}
where
\begin{align}
 \cF_{ab}{}^c=f_{ab}{}^c\,,\quad 
 \cF_{abc}=0\,,\quad
 \cF_{a}{}^{bc}=\tilde{f}^{bc}{}_a\,,\quad 
 \cF_{a}{}^b{}_{c} = -f_{ac}{}^b \,,\quad 
 \cF^{ab}{}_c = \tilde{f}^{ab}{}_c \,,\quad 
 \cF^{abc}=0\,.
\label{eq:cF-Drinfeld}
\end{align}
For simplicity, we here suppose that the structure constant of the dual algebra is unimodular $\tilde{f}^{ba}{}_b=0$\,.\footnote{See \cite{1810.11446,1903.12175} for the PL $T$-duality for non-unimodular cases.} 
Then, the equations of motion of DFT for a general PL $T$-dualizable background \eqref{eq:PL-dualizable-BG} reduce to the following algebraic equations:
\begin{align}
\begin{split}
 &\frac{1}{12}\,\cF_{ABC}\,\cF_{DEF} \,\bigl(3\,\hat{\cH}^{AD}\,\eta^{BE}\,\eta^{CF}- \hat{\cH}^{AD}\,\hat{\cH}^{BE}\,\hat{\cH}^{CF}\bigr) = 0 \,,
\\
 &\frac{1}{2}\,\bigl(\eta^{CE}\,\eta^{DF} - \hat{\cH}^{CE}\,\hat{\cH}^{DF} \bigr)\,\hat{\cH}^{G[A}\,\cF_{CD}{}^{B]}\,\cF_{EFG} = 0 \,.
\end{split}
\label{eq:DFT-eom}
\end{align}
They are manifestly covariant under constant $\OO(D,D)$ transformations\footnote{The matrix $C_A{}^B$ should be chosen such that the structure constant $\cF'_{AB}{}^C$ has the form of the Drinfel'd double \eqref{eq:cF-Drinfeld}. If this is not obeyed, we do not have a systematic way to construct $E'_A{}^M$ satisfying \eqref{eq:E-F-prime}.}
\begin{align}
 \hat{\cH}'_{AB} = C_A{}^C\,C_B{}^D\,\hat{\cH}_{CD}\,,\qquad 
 \hat{d}'= \hat{d} \,,\qquad 
 \cF'_{ABC} = C_A{}^E\,C_B{}^F\,C_C{}^G\,\cF_{EFG}\,.
\label{eq:PL-transf}
\end{align}
This is the PL $T$-duality, which extends the Abelian $T$-duality \eqref{eq:Ableian-T-dual}.\footnote{To be more precise, the PL $T$-duality is a particular transformation $(C_A{}^B)=\bigl(\begin{smallmatrix} 0 & \bm{1}_D \\ \bm{1}_D & 0\end{smallmatrix} \bigr)$\,. In particular, when $\tilde{f}^{ab}{}_c=0$\,, it is called the non-Abelian $T$-duality. A general $\OO(D,D)$ transformation is called the PL $T$-plurality transformation \cite{hep-th:0205245}, but in this paper, we denote an arbitrary $\OO(D,D)$ transformation as the PL $T$-duality.}

In DFT, the tensor $\cF_{AB}{}^C$ is called the generalized flux, and it is generally defined as
\begin{align}
 \gLie_{E_A} E_B{}^M = - \cF_{AB}{}^C\,E_C{}^M \,,
\label{eq:E-F}
\end{align}
Here, the generalized frame fields $E_A=(E_A{}^M)$ correspond to the inverse of the twist matrix $E_M{}^A$ given in \eqref{eq:twist-E}. 
Even in general spacetimes where the generalized flux $\cF_{AB}{}^C$ is not constant, the equations of motion of DFT can be expressed using $\cF_{AB}{}^C$\,. 
However, they contain the derivative of the generalized flux $\partial_M \cF_{AB}{}^C$ and are much more complicated than \eqref{eq:DFT-eom}. 
In such general cases, the $\OO(D,D)$ transformation \eqref{eq:PL-transf} is not a symmetry of DFT. 
Therefore, the constancy of the generalized flux $\cF_{AB}{}^C$ is crucial for the PL $T$-duality. 

\paragraph{Dual geometry:}
Under a PL $T$-duality, the structure constant $\cF_{ABC}$ is mapped to another one $\cF'_{ABC}$\,. 
It is associated with a new set of the generalized frame fields $E'_A{}^M$ satisfying
\begin{align}
 \gLie_{E'_A} E'_B{}^M = - \cF'_{AB}{}^C\,E'_C{}^M \,.
\label{eq:E-F-prime}
\end{align}
Then, by using the new twist matrix $E'_A{}^M$ and the relation \eqref{eq:PL-dualizable-BG}, the metric, $B$-field, and the dilaton in the dual geometry are obtained as
\begin{align}
 \cH'_{MN} = E'_M{}^A\,E'_N{}^B\, \hat{\cH}'_{AB} \,,\qquad \Exp{-2d'} = \Exp{-2\hat{d}'}\abs{\det(\ell'^a_m)}\,. 
\end{align}

A possible problem in the PL $T$-duality is the explicit construction of $E'_A{}^M$ satisfying \eqref{eq:E-F-prime}. 
However, it is not a problem if we use the algebra of the Drinfel'd double. 
Under the PL $T$-duality \eqref{eq:PL-transf}, the generators are also redefined as $T'_A = C_A{}^B\,T_B$\,, and they satisfy
\begin{align}
 [T'_A,\,T'_B] = \cF'_{AB}{}^C\,T_C \,,\qquad \langle T'_A,\,T'_B\rangle = \eta_{AB} \,.
\end{align}
Then, by decomposing the new generators as $(T'_A)=(T'_a,\,\tilde{T}'^a)$, we obtain a new physical subalgebra $\mathfrak{g}'$ spanned by $T'_a$\,. 
By parameterizing a group element as before, such as $g'\equiv \Exp{x'^a\,T'_a}$, we can again obtain the matrices $\Pi'^{ab}$, $\ell'^a_m$, and $r'^a_m$\,. 
Then, the new generalized frame fields
\begin{align}
 (E'_A{}^M) \equiv \begin{pmatrix} e'^m_a & 0 \\ \Pi'^{a c}\,e'^m_c & r'^a_m \end{pmatrix}
\label{eq:E-prime-DFT}
\end{align}
satisfy the expected relation \eqref{eq:E-F-prime}. 
In this manner, we can systematically construct the dual geometry, and the construction method of $E_A{}^M$ is important for the PL $T$-duality. 

\paragraph{A short summary:}
As shortly reviewed in this section, the PL $T$-duality is a constant $\OO(D,D)$ rotation of the indices $A,B$\,. 
We can perform this duality when the generalized frame fields satisfy the relation \eqref{eq:E-F} by using the structure constant $\cF_{AB}{}^C$ of a Drinfel'd double. 
The algebra of the Drinfel'd double provides a systematic way to construct the generalized frame fields satisfying \eqref{eq:E-F-prime}, and the dual geometry can be explicitly constructed. 

In the next section, we introduce an extension of the Drinfel'd double and explain a systematic way to construct a generalized frame fields satisfying
\begin{align}
 \gLie_{E_A} E_B{}^I = - \cF_{AB}{}^C\,E_C{}^I \quad (\cF_{AB}{}^C:\text{ constant})
\label{eq:EFT-gLie}
\end{align}
by means of the generalized Lie derivative in EFT. 

\section{Leibniz algebra based on $U$-duality}
\label{sec:algebra}

Here, we propose a Leibniz algebra $\cE_n$ by using the generalized Lie derivative in the $E_n$ EFT. 
For this purpose, let us begin with a quick introduction to EFT. 

\paragraph{Basic definitions in EFT:}
As we have explained in the introduction, in EFT, we introduce an exceptional space with dimension $D_n$\,. 
When we adopt the M-theory picture, we decompose the generalized coordinates $x^I$ ($I=1,\dotsc,D_n$) as \cite{hep-th:0307098}
\begin{align}
 (x^I) = \bigl(x^i,\,\tfrac{y_{i_1i_2}}{\sqrt{2!}},\,\tfrac{y_{i_1\cdots i_5}}{\sqrt{5!}}\,,\cdots \bigr)\qquad (i= 1,\dotsc,n)\,,
\end{align}
where the multiple indices are totally antisymmetric and the numerical factors are introduced for convenience. 
The ellipses are not necessary as far as we consider the cases $n\leq 6$. 
The supergravity fields such as the metric and gauge potentials are contained in the generalized metric $\cM_{IJ}$, which extends the one $\cH_{MN}$ in DFT. 
The fields are formally defined on the exceptional space, but the extension of the section condition \eqref{eq:SC-DFT} again restricts the coordinate dependence. 
In order to reproduce M-theory, we choose $x^i$ as the physical coordinates and any more coordinate dependence is not allowed by the section condition. 
Thus, in the following discussion, we eliminate the coordinate dependence on the dual coordinates:
\begin{align}
 \frac{\partial}{\partial y_{i_1i_2}} = 0\,,\qquad 
 \frac{\partial}{\partial y_{i_1\cdots i_5}} = 0\,,\qquad \cdots\,.
\label{eq:M-section}
\end{align}

Similar to DFT, the generalized Lie derivative in EFT is defined as \cite{1208.5884}
\begin{align}
 \gLie_V W^I \equiv V^J\,\partial_J W^I - W^J\,\partial_J V^I + Y^{IJ}_{KL}\, \partial_J V^K\,W^L\,,
\end{align}
where $Y^{IJ}_{KL}$ is an invariant tensor satisfying $\gLie_V Y^{IJ}_{KL}=0$\,. 
For our purpose, it is enough to know the expression under the situation where all fields depend only on the physical coordinates $x^i$\,. 
In that case, the generalized Lie derivative is expressed as (see \cite{1708.06342} for our convention)
\begin{align}
 \bigl(\gLie_V W^I\bigr) = 
 \begin{pmatrix}
 \Lie_v w^i \\
 \frac{(\Lie_v w_2 - \iota_w \rmd v_2)_{i_1i_2}}{\sqrt{2!}} \\
 \frac{(\Lie_v w_5 + \rmd v_2\wedge w_2 - \iota_w\rmd v_5)_{i_1\cdots i_5}}{\sqrt{5!}} \\
 \vdots
 \end{pmatrix},
\label{eq:gen-Lie-EFT}
\end{align}
where the two arbitrary generalized vectors $V^I$ and $W^I$ are parameterized as
\begin{align}
 (V^I) = 
 \begin{pmatrix}
 v^i \\
 \frac{v_{i_1i_2}}{\sqrt{2!}} \\
 \frac{v_{i_1\cdots i_5}}{\sqrt{5!}} \\
 \vdots
 \end{pmatrix},\qquad 
 (W^I) = 
 \begin{pmatrix}
 w^i \\
 \frac{w_{i_1i_2}}{\sqrt{2!}} \\
 \frac{w_{i_1\cdots i_5}}{\sqrt{5!}} \\
 \vdots
 \end{pmatrix},
\end{align}
and we have defined $v_p \equiv \frac{1}{p!}\,v_{i_1\cdots i_p}\,\rmd x^{i_1}\wedge\cdots\wedge\rmd x^{i_p}$ and similar for $w_p$\,. 
We note that the expression \eqref{eq:gen-Lie-EFT} coincides with the Dorfman derivative in generalized geometry \cite{1112.3989,1212.1586}. 

In order to simplify our discussion, we restrict our attention to $n\leq 4$\,. 
Then, terms with five (or more) antisymmetrized indices identically vanish (e.g.~$v_{i_1\cdots i_5}=0$) and the above generalized vectors reduce to
\begin{align}
 (V^I) = 
 \begin{pmatrix}
 v^i \\
 \frac{v_{i_1i_2}}{\sqrt{2!}} 
 \end{pmatrix},\qquad 
 (W^I) = 
 \begin{pmatrix}
 w^i \\
 \frac{w_{i_1i_2}}{\sqrt{2!}} 
 \end{pmatrix},\qquad
 \bigl(\gLie_V W^I\bigr) = 
 \begin{pmatrix}
 \Lie_v w^i \\
 \frac{(\Lie_v w_2 - \iota_w \rmd v_2)_{i_1i_2}}{\sqrt{2!}} \end{pmatrix}.
\label{eq:V-W}
\end{align}

\paragraph{Generalized frame fields in EFT:}
In order to consider the relation \eqref{eq:E-F} in EFT, let us introduce certain generalized frame fields $E_A{}^I$ in EFT. 
By considering the analogy with the DFT case \eqref{eq:E-prime-DFT}, we consider the following parameterization:
\begin{align}
 (E_A{}^I) &\equiv \begin{pmatrix} E_a{}^I \\ \frac{E^{a_1a_2}{}^I}{\sqrt{2!}} \end{pmatrix}
 \equiv \begin{pmatrix} \delta_a^b & 0 \\ -\tfrac{\Pi^{a_1a_2 b}}{\sqrt{2!}} & \delta^{a_1a_2}_{b_1b_2} \end{pmatrix}
 \begin{pmatrix} e_b^i & 0 \\ 0 & r^{[b_1}_{[i_1}\,r^{b_2]}_{i_2]} \end{pmatrix}
\nn\\
 &=\begin{pmatrix} e_a^i & 0 \\ -\tfrac{\Pi^{a_1a_2 b}\,e_b^i}{\sqrt{2!}} & r^{[a_1}_{[i_1}\,r^{a_2]}_{i_2]} \end{pmatrix} ,
\label{eq:gen-frame}
\end{align}
where $\delta^{a_1\cdots a_n}_{b_1\cdots b_n}\equiv \delta^{[a_1}_{[b_1}\cdots \delta^{a_n]}_{b_n]}$ and $\Pi^{a_1a_2a_3}=\Pi^{[a_1a_2a_3]}$, and $e_a^m$ (or $r^a_m$) is a certain right-invariant vector (or 1-form) satisfying
\begin{align}
 \Lie_{e_a} e_b^m = - f_{ab}{}^c\,e_c^m\,,\qquad 
 \rmd r^a = \frac{1}{2}\,f_{bc}{}^a\, r^b\wedge r^c \,,\qquad 
 \iota_{e_a} r^b = \delta_a^b \,.
\end{align}
Then, using \eqref{eq:V-W}, we can compute the generalized Lie derivative as follows:
\begin{align}
 \gLie_{E_A} E_B{}^I = - X_{AB}{}^C\, E_C{}^I\,,
\end{align}
where
\begin{align}
 X_{ab}{}^c &= f_{ab}{}^c\,,
\\
 X_{abc_1c_2}&=0\,,
\\
 X_{a}{}^{b_1b_2}{}^c &= D_a\Pi^{b_1b_2 c} - 3\,f_{ad}{}^{[c} \, \Pi^{b_1b_2]d} \,,
\\
 X_{a}{}^{b_1b_2}{}_{c_1c_2} &= 4\,f_{ad}{}^{e}\, \delta^{b_1b_2}_{ef}\,\delta_{c_1c_2}^{fd}\,,
\\
 X^{a_1a_2}{}_b{}^c &= -\bigl(D_b\Pi^{a_1a_2 c} - 3\,f_{bd}{}^{[c} \, \Pi^{a_1a_2]d} - f_{d_1d_2}{}^{[a_1}\,\delta^{a_2]}_{b}\,\Pi^{d_1d_2 c}\bigr) \,,
\\
 X^{a_1a_2}{}_{b c_1c_2} &= 6\, f_{[bc_1}{}^{[a_1}\,\delta^{a_2]}_{c_2]} \,,
\\
 X^{a_1a_2b_1b_2c} &= - \Pi^{a_1a_2d}\, D_d \Pi^{b_1b_2c}
 + 3\,\Pi^{[b_1b_2|d}\, D_d \Pi^{a_1a_2|c]} 
\nn\\
 &\quad + 2\, f_{gh}{}^{[a_1}\, \Pi^{a_2]cg}\, \Pi^{b_1b_2 h}
 - 3\, f_{gh}{}^{[b_1}\, \Pi^{b_2c] g}\, \Pi^{a_1a_2 h} 
 - f_{gh}{}^{[a_1}\,\Pi^{a_2]b_1b_2}\, \Pi^{gh c} \,,
\\
 X^{a_1a_2b_1b_2}{}_{c_1c_2} &= -\bigl(4\, D_d\Pi^{a_1a_2[b_1}\, \delta^{b_2]d}_{c_1c_2} 
 + 4\, \Pi^{a_1a_2 d}\, f_{de}{}^{[b_1}\,\delta^{b_2]e}_{c_1c_2} 
 - 4\, \Pi^{b_1b_2 d}\, f_{de}{}^{[a_1}\,\delta^{a_2]e}_{c_1c_2} 
\nn\\
 &\qquad + 2\, f_{c_1c_2}{}^{[a_1}\,\Pi^{a_2]b_1b_2}\bigr) \,,
\end{align}
and $D_a \equiv e_a^i\,\partial_i$\,. 
In general, the generalized flux $X_{AB}{}^C$ is not constant.\footnote{See \cite{1412.0635,1901.07775} for computation of the generalized flux in the $\SL(5)$ EFT in a more general setup.}

Unlike the DFT case, the first two indices are not antisymmetric---$X_{AB}{}^C\neq X_{[AB]}{}^C$---and even if we find a certain situation where $X_{AB}{}^C$ is constant, the algebra is not a Lie algebra. 
Accordingly, in the following, we investigate a Leibniz algebra satisfying
\begin{align}
 T_A \circ T_B = \cF_{AB}{}^C \, T_C \,.
\label{eq:TA-TB-X}
\end{align}
Here, $T_A \circ T_B\neq - T_B \circ T_A$ but the Leibniz identity,
\begin{align}
 X\circ(Y\circ Z) - Y\circ (X\circ Z) = (X\circ Y)\circ Z \,,
\end{align}
is satisfied similar to the case of the generalized Lie derivative,
\begin{align}
 \gLie_{V_1} \gLie_{V_2} W^I - \gLie_{V_2}\gLie_{V_1} W^I = \gLie_{\gLie_{V_1}V_2} W^I \,. 
\end{align}

\paragraph{Construction of the algebra $\cE_n$:}
Here we take a heuristic approach to find the Leibniz algebra $\cE_n$\,. 
First, we suppose that the generalized flux $X_{AB}{}^C$ is constant, and assume that there exists an algebra \eqref{eq:TA-TB-X} with $\cF_{AB}{}^C = X_{AB}{}^C$\,. 
Secondly, we assume that $\Pi^{a_1a_2a_3}=0$ at a certain point $x^a=0$\,, which corresponds to \eqref{eq:Pi=0}. 
This shows that the so-called $R$-flux $X^{a_1a_2b_1b_2c}$ vanishes. 
At least when $e_a^i=\delta_a^i$, this is precisely a condition for $\Pi^{a_1a_2a_3}$ to be a Nambu--Poisson tensor \cite{hep-th:9301111}, and the condition $X^{a_1a_2b_1b_2c}=0$ will be understood as a natural generalization of the definition of the Nambu--Poisson tensor. 
In the case of the Drinfel'd double, the bi-vector $\Pi^{ab}$ has been a Poisson tensor, and in our setup the Poisson tensor is naturally extended to the Nambu--Poisson tensor. 

Under these assumptions, the generalized flux $X_{AB}{}^C$ at the point $x^a=0$ reduces to
\begin{align}
 X_{ab}{}^c &= f_{ab}{}^c\,, 
\\
 X_{abc_1c_2}&=0\,,
\\
 X_{a}{}^{b_1b_2}{}^c &= D_a\Pi^{b_1b_2 c} = X_{a}{}^{[b_1b_2}{}^{c]}\,,
\\
 X_{a}{}^{b_1b_2}{}_{c_1c_2} &= 4\,f_{ad}{}^{e}\, \delta^{b_1b_2}_{ef}\,\delta_{c_1c_2}^{fd}\,,
\\
 X^{a_1a_2}{}_b{}^c &= -D_b\Pi^{a_1a_2 c} = - X_{b}{}^{a_1a_2}{}^c\,,
\\
 X^{a_1a_2}{}_{b c_1c_2} &= 6\, f_{[bc_1}{}^{[a_1}\,\delta^{a_2]}_{c_2]} \,,
\\
 X^{a_1a_2b_1b_2c} &= 0 \,,
\\
 X^{a_1a_2b_1b_2}{}_{c_1c_2} &= -4\, D_d\Pi^{a_1a_2[b_1}\, \delta^{b_2]d}_{c_1c_2} = -4\,X_{d}{}^{a_1a_2}{}^{[b_1}\, \delta^{b_2]d}_{c_1c_2} \,.
\end{align}
This prompts us to define a new Leibniz algebra $\cE_n$ as
\begin{align}
\begin{split}
 T_a\circ T_b &= f_{ab}{}^c\,T_c \,,
\\
 T_a\circ T^{b_1b_2} &= f_a{}^{b_1b_2c}\,T_c + 2\,f_{ac}{}^{[b_1}\, T^{b_2]c} \,,
\\
 T^{a_1a_2}\circ T_b &= -f_b{}^{a_1a_2 c}\,T_c + 3\,f_{[c_1c_2}{}^{[a_1}\,\delta^{a_2]}_{b]} \,T^{c_1c_2}\,,
\\
 T^{a_1a_2}\circ T^{b_1b_2} &= -2\, f_d{}^{a_1a_2[b_1}\, T^{b_2]d}\,,
\end{split}
\end{align}
where $f_{ab}{}^c=f_{[ab]}{}^c$ and $f_a{}^{b_1b_2b_3}=f_a{}^{[b_1b_2b_3]}$\,. 
The Leibniz identity
\begin{align}
 X\circ(Y\circ Z) = (X\circ Y)\circ Z + Y\circ (X\circ Z)
\end{align}
for the generators $T_a$ and $T^{a_1a_2}$ requires the following relations:\footnote{An additional relation $0 = 3\,f_{[d_1d_2}{}^{[a_1}\,\delta^{a_2]}_{e]}\,f_c{}^{e b_1b_2} + 4\,f_{ef}{}^{[a_1}\,f_c{}^{a_2]e[b_1}\,\delta^{b_2]f}_{d_1d_2}$ was given in the previous versions. However, for $n\leq 4$, this follows from \eqref{eq:Leibniz-3}. Indeed, under \eqref{eq:Leibniz-3} or $f_{d_1d_2}{}^{a}\, f_c{}^{d_1d_2 e}=0$, the right-hand side of this relation is equal to $5\,\bigl(f_{e_1e_2}{}^{a_1}\,f_c{}^{[a_2b_1b_2}\,\delta^{e_1e_2]}_{d_1d_2} -f_{e_1e_2}{}^{a_2}\,f_c{}^{[a_1b_1b_2}\,\delta^{e_1e_2]}_{d_1d_2}\bigr)$ which trivially vanishes for $n\leq 4$.}
\begin{align}
 0&=f_{[ab}{}^{e}\,f_{c]e}{}^d\,,
\\
 0&=f_{bc}{}^e\,f_{e}{}^{a_1a_2 d} + 6\, f_{e[b}{}^{[d}\, f_{c]}{}^{a_1a_2]e} \,,
\\
 0&=f_{d_1d_2}{}^{[a_1}\, \delta^{a_2]}_b\, f_c{}^{d_1d_2 e} \,,
\label{eq:Leibniz-3}
\\
 0&=f_{c}{}^{e a_1a_2}\,f_e{}^{db_1b_2} - 3\, f_c{}^{e[b_1b_2}\, f_e{}^{d] a_1a_2} \,.
\end{align}

\paragraph{A bilinear form:}
We also introduce the bilinear form, which extends the bilinear form $\langle \cdot,\,\cdot \rangle$ of the Drinfel'd double. 
A natural extension of the bilinear form has been known in EFT,\footnote{This bilinear form has also been studied in a mathematical literature \cite{math:0204310}.}
\begin{align}
 \langle V,\,W \rangle_{\cK} \equiv \eta_{IJ;\,\cK}\,V^I\,W^J\,,
\end{align}
where $\eta_{IJ;\,\cK}$ connects a product of two $R_1$-representations and another representation, called the $R_2$-representation (see for example \cite{1208.5884}), whose dimension $d_n$ is given in Table \ref{table:En}. 
Namely, the additional index $\cK$ appended to the bilinear form transforms in the $R_2$-representation. 
This index can be decomposed as (see \cite{1708.06342} for the explicit form of $\eta_{IJ;\,\cK}$)
\begin{align}
 (\eta_{IJ;\,\cK}) = \bigl(\eta_{IJ;\,k},\,\tfrac{\eta_{IJ;\,k_1\cdots k_4}}{\sqrt{4!}},\cdots\bigr)\,,
\end{align}
and in our case $n\leq 4$\,, it is enough to consider the first two components,
\begin{align}
 \langle \cdot,\,\cdot \rangle_\cK = \Bigl(\langle \cdot,\,\cdot \rangle_k,\,\tfrac{\langle \cdot,\,\cdot \rangle_{k_1\cdots k_4}}{\sqrt{4!}} \Bigr)\,.
\end{align}
The bilinear form takes the form
\begin{align}
 \langle V,\,W \rangle_k = \bigl(\iota_w v_2 + \iota_v w_2\bigr)_k\,, \qquad
 \langle V,\,W \rangle_{k_1\cdots k_4} = \bigl(v_2\wedge w_2\bigr)_{k_1\cdots k_4}\,,
\end{align}
for two arbitrary vectors $V^I$ and $W^I$ parameterized as \eqref{eq:V-W}. 
Under an arbitrary $E_n$ $U$-duality transformation $\Lambda$, the tensor $\eta_{IJ;\,\cK}$ behaves as
\begin{align}
 \Lambda_I{}^{L_1}\,\Lambda_J{}^{L_2}\,\Lambda_{\cK}{}^{\cL}\,\eta_{L_1L_2;\,\cL} = \eta_{IJ;\,\cK}\,,
\end{align}
where $\Lambda_I{}^J$ and $\Lambda_{\cI}{}^{\cJ}$ denote the same $E_n$ transformation in the $R_1$- and $R_2$-representation, respectively. 

Now, we introduce a matrix,
\begin{align}
 (E_\cA{}^\cI) \equiv \begin{pmatrix} \delta_a^b & -\frac{4\,\delta_a^{[b_1} \Pi^{b_2b_3b_4]}}{\sqrt{4!}} \\ 0 & \delta_{a_1\cdots a_4}^{b_1\cdots b_4} \end{pmatrix}\begin{pmatrix} e_b^i & 0 \\ 0 & e_{[b_1}^{[i_1}\cdots e_{b_4]}^{i_4]} 
\end{pmatrix},
\end{align}
which satisfies
\begin{align}
 E_A{}^I\,E_B{}^J\,E_\cC{}^\cK \,\eta_{IJ;\,\cK} = \eta_{AB;\,\cC} \,,
\end{align}
and redefine the bilinear form as
\begin{align}
 \langle \cdot,\,\cdot \rangle_\cA \equiv \Bigl(\langle \cdot,\,\cdot \rangle_a,\,\tfrac{\langle \cdot,\,\cdot \rangle_{a_1\cdots a_4}}{\sqrt{4!}} \Bigr) \equiv E_{\cA}{}^\cI\, \langle \cdot,\, \cdot \rangle_\cI \,. 
\end{align}
Then, the bilinear form for the generalized frame fields \eqref{eq:gen-frame} becomes
\begin{align}
 \langle E_a,\, E^{b_1b_2}\rangle_c = 2!\,\delta^{b_1b_2}_{ca} \,,\qquad 
 \langle E^{a_1a_2},\, E^{b_1b_2}\rangle_{c_1\cdots c_4} = 4!\,\delta^{a_1a_2b_1b_2}_{c_1\cdots c_4} \,.
\end{align}
Identifying the generalized frame fields $E_A$ with the $\cE_n$ generator $T_A$\,, we define the following bilinear form for the generators:
\begin{align}
 \langle T_a,\, T^{b_1b_2}\rangle_c = 2!\,\delta^{b_1b_2}_{ac} \,,\qquad 
 \langle T^{a_1a_2},\, T^{b_1b_2}\rangle_{c_1\cdots c_4} = 4!\,\delta^{a_1a_2b_1b_2}_{c_1\cdots c_4}\,.
\end{align}

We note that the subalgebra spanned by $\{T_a\}$ is maximally isotropic for the bilinear form. 
In fact, the isotropicity shows that the subalgebra is a Lie algebra $T_a\circ T_b = [T_a,\,T_b]$\,, where
\begin{align}
 [T_A,\,T_B] \equiv \frac{1}{2} \,\bigl(T_A \circ T_B - T_B\circ T_A\bigr)\,. 
\end{align}
This can be understood from the explicit form of the generalized Lie derivative \eqref{eq:V-W}, namely,
\begin{align}
 \bigl(\gLie_V W^I\bigr) = 
 \begin{pmatrix}
 [v,\, w]^i \\
 \frac{[\Lie_v w_2 - \Lie_w v_2 + \rmd (\iota_w v_2)]_{i_1i_2}}{\sqrt{2!}} \end{pmatrix}.
\end{align}
When $V^I$ and $W^I$ satisfy $\langle V,\,W \rangle_\cA=0$\,, we have
\begin{align}
 \iota_w v_2 = \frac{1}{2}\,\bigl(\iota_w v_2 - \iota_v w_2 \bigr)\,,
\end{align}
and the generalized Lie derivative satisfies $\gLie_V W^I = - \gLie_W V^I$\,. 
Accordingly, for a set of the generalized frame fields $\{E_a\}$ forming an isotropic subalgebra, we have
\begin{align}
 \gLie_{E_a} E_b{}^I = \frac{1}{2}\,\bigl(\gLie_{E_a} E_b{}^I - \gLie_{E_b} E_a{}^I\bigr) = - X_{[ab]}{}^c\,E_c{}^I\,,
\end{align}
and the subalgebra is a Lie algebra. 
This property plays an important role when we explicitly construct the generalized frame fields. 

\paragraph{$E_n$ generators:}
For the sake of clarity, let us explain our convention for the $E_n$ generators. 
We decompose the $E_n$ generators $\{t_{\hat{\alpha}}\}$ $(\hat{\alpha}=1,\dotsc,\dim E_n)$ for $n\leq 4$ as \cite{hep-th:0104081}
\begin{align}
 (t_{\hat{\alpha}}) \equiv \bigl( K^c{}_d\,,\ \tfrac{R^{c_1c_2c_3}}{\sqrt{3!}}\,,\ \tfrac{R_{c_1c_2c_3}}{\sqrt{3!}}\bigr)\,.
\label{eq:En-generators}
\end{align}
Their matrix representations $(t_{\hat{\alpha}})_A{}^B$ in the $R_1$-representation are given as follows:\footnote{The second term in the $\GL(n)$ generator $(K^a{}_b)_A{}^B$, which is proportional to the identity matrix $\delta_A^B$, is necessary for the commutator $[R^{c_1c_2c_3},\,R_{d_1d_2d_3}]$ to be expanded by the generator $K^a{}_b$\,. We also note that the two $\GL(n)$ generators $\tilde{K}^a{}_b$ and $K^a{}_b$ provide two notions of the weight, ``effective weight'' and ``weight'' (see \cite{1312.0614}). The diffeomorphism parameters and $E_A{}^I$ have ``effective weight'' $0$ and ``weight'' $\frac{1}{9-n}$\,.\label{footnote:GL(n)}}
\begin{align}
\begin{split}
\begin{alignedat}{2}
 (K^c{}_d)_A{}^B &\equiv (\tilde{K}^c{}_d)_A{}^B + \frac{\delta_c^d}{9-n}\,\delta_A^B\,,\quad&
 (\tilde{K}^c{}_d)_A{}^B &\equiv \begin{pmatrix} \delta^c_a \delta^b_d & 0 \\ 0 & -2\,\delta_{de}^{a_1a_2}\,\delta_{b_1b_2}^{ce} \end{pmatrix} \,, 
\\
 (R^{c_1c_2c_3})_A{}^B &\equiv \begin{pmatrix} 0 & \frac{3!\,\delta_{a b_1b_2}^{c_1c_2c_3}}{\sqrt{2!}} \\ 0 & 0 \end{pmatrix} , \quad&
 (R_{c_1c_2c_3})_A{}^B &\equiv \begin{pmatrix} 0 & 0 \\ \frac{3!\,\delta^{b a_1a_2}_{c_1c_2c_3}}{\sqrt{2!}} & 0 \end{pmatrix}.
\end{alignedat}\end{split}
\label{eq:En-l1}
\end{align}
The matrix representations $(t_{\hat{\alpha}})_{\cA}{}^{\cB}$ in the $R_2$-representation are
\begin{align}
\begin{split}
\begin{alignedat}{2}
 (K^c{}_d)_{\cA}{}^{\cB} &\equiv (\tilde{K}^c{}_d)_{\cA}{}^{\cB} - \frac{2\,\delta^c_d}{9-n}\, \delta_{\cA}^{\cB}\,,\quad &
 (\tilde{K}^c{}_d)_{\cA}{}^{\cB} &\equiv \begin{pmatrix}
 \delta^c_a\,\delta_d^b & 0 \\
 0 & 4\,\delta^{c e_1e_2e_3}_{a_1\cdots a_4}\, \delta^{b_1\cdots b_4}_{d e_1e_2e_3} \end{pmatrix} , 
\\
 (R^{c_1c_2c_3})_{\cA}{}^{\cB} &\equiv 
 \begin{pmatrix} 0 & 0 \\ \frac{4!\,\delta^{b c_1c_2c_3}_{a_1\cdots a_4}}{\sqrt{4!}} & 0 \end{pmatrix}, \quad&
 (R_{c_1c_2c_3})_{\cA}{}^{\cB} &\equiv \begin{pmatrix} 0 & \frac{4!\,\delta_{a s_1s_2s_3}^{b_1\cdots b_4}}{\sqrt{4!}} \\ 0 & 0 \end{pmatrix}. 
\end{alignedat}\end{split}
\end{align}

Now, let us rewrite the $\cE_n$ algebra. 
If we express the algebra as
\begin{align}
 T_C\circ T_A = (T_C)_A{}^B\,T_B\,,
\end{align}
the matrices $(T_C)_A{}^B$ are given by
\begin{align}
 (T_c)_A{}^B &= \begin{pmatrix} f_{ca}{}^b & 0 \\ \frac{f_c{}^{a_1a_2b}}{\sqrt{2!}} & -2\,f_{c[b_1}{}^{[a_1}\,\delta^{a_2]}_{b_2]}\end{pmatrix} \,,
\label{eq:Tc-matrix}
\\
 (T^{c_1c_2})_A{}^B &= \begin{pmatrix} -f_a{}^{c_1c_2 b} & \frac{6\,f_{[b_1b_2}{}^{[c_1}\,\delta^{c_2]}_{a]}}{\sqrt{2!}} \\ 0 & -2\, f_d{}^{c_1c_2[a_1}\,\delta^{a_2]d}_{b_1b_2} \end{pmatrix} \,.
\end{align}
They can be expressed as
\begin{align}
 (T_c)_A{}^B &= f_{cd}{}^e\,(\tilde{K}^d{}_e)_A{}^B + \frac{1}{3!}\, f_c{}^{d_1d_2d_3} \, (R_{d_1d_2d_2})_A{}^B\,,
\\
 (T^{c_1c_2})_A{}^B &= -f_d{}^{c_1c_2 e}\,(\tilde{K}^d{}_e)_A{}^B + f_{[d_1d_2}{}^{[c_1}\,\delta^{c_2]}_{d_3]}\,(R^{d_1d_2d_3})_A{}^B\,.
\end{align}
In general, they are not exactly $E_n$ $U$-duality transformations, because $\tilde{K}^a{}_b$ is not an $E_n$ generator. 
Thus, suggested by \cite{1112.3989,1302.5419}, we introduce an additional generator $(t_0)_A{}^B\equiv -\delta_A^B$ for $\lR^+$ \cite{1302.5419}, which is associated with the scaling of the density, and express the algebra as
\begin{align}
 (T_c)_A{}^B &= f_{cd}{}^e\,(K^d{}_e)_A{}^B + \frac{1}{3!}\, f_c{}^{d_1d_2d_3} \, (R_{d_1d_2d_2})_A{}^B + \frac{f_{cd}{}^d}{9-n}\,(t_0)_A{}^B\,,
\\
 (T^{c_1c_2})_A{}^B &= -f_d{}^{c_1c_2 e}\,(K^d{}_e)_A{}^B + f_{[d_1d_2}{}^{[c_1}\,\delta^{c_2]}_{d_3]}\,(R^{d_1d_2d_3})_A{}^B - \frac{f_d{}^{c_1c_2 d}}{9-n}\,(t_0)_A{}^B\,.
\end{align}
The coefficient in the last term of each line is due to the generalized frame fields $E_A{}^I$ having the density weight $\frac{1}{9-n}$ (see footnote \ref{footnote:GL(n)}). 
Then, the $\cE_n$ algebra can be also expressed as
\begin{align}
 T_A \circ T_B = \bigl[\Theta_A^{\hat{\alpha}}\, (t_{\hat{\alpha}})_B{}^C + \theta_A\,(t_0)_B{}^C\bigr]\,T_C\,,
\end{align}
where $\Theta_A^{\hat{\alpha}}$ and $\theta_A$ are constants. 
If we decompose the index $\hat{\alpha}$ as
\begin{align}
 (\Theta_A^{\hat{\alpha}}) = \bigl([\Theta_A]_a{}^b\,,\ \tfrac{[\Theta_A]_{a_1a_2a_3}}{\sqrt{3!}}\,,\ \tfrac{[\Theta_A]^{a_1a_2a_3}}{\sqrt{3!}} \bigr) \,,
\end{align}
their components are
\begin{align}
\begin{split}
 [\Theta_a]_b{}^c &= f_{ab}{}^c\,,\qquad 
 [\Theta_a]_{c_1c_2c_3} = 0\,, \qquad
 [\Theta_a]^{c_1c_2c_3} = f_a{}^{c_1c_2c_3}\,,
\\
 [\Theta^{a_1a_2}]_b{}^c &= - f_b{}^{a_1a_2 c} \,,\qquad 
 [\Theta^{a_1a_2}]_{c_1c_2c_3} = 3!\,f_{[c_1c_2}{}^{[a_1}\,\delta^{a_2]}_{c_3]}\,,\qquad
 [\Theta^{a_1a_2}]^{c_1c_2c_3} = 0\,, 
\\
 \theta_a &= \frac{f_{ad}{}^d}{9-n}\,,\qquad
 \theta^{a_1a_2} = \frac{f_d{}^{c_1c_2 d}}{9-n}\,.
\end{split}
\label{eq:En-assumption}
\end{align}

Then, we can easily obtain the matrices $(T_C)_\cA{}^\cB$ in the $R_2$-representation as follows:
\begin{align}
 (T_c)_\cA{}^\cB &= f_{cd}{}^e\,(K^d{}_e)_{\cA}{}^{\cB} + \frac{1}{3!}\, f_c{}^{d_1d_2d_3} \, (R_{d_1d_2d_2})_{\cA}{}^{\cB} + \frac{f_{cd}{}^d}{9-n}\,(t_0)_\cA{}^\cB\,,
\nn\\
 &= \begin{pmatrix} f_{ca}{}^b & \frac{4\,\delta_a^{[b_1}\,f_c{}^{b_2b_3b_4]}}{\sqrt{4!}} \\ 0 & 4\,f_{c[a_1}{}^{[b_1}\,\delta^{b_2b_3b_4]}_{a_2a_3a_4]} \end{pmatrix} ,
\\
 (T^{c_1c_2})_\cA{}^\cB &= - f_d{}^{c_1c_2 e}\,(K^d{}_e)_{\cA}{}^{\cB} + f_{[d_1d_2}{}^{[c_1}\,\delta^{c_2]}_{d_3]}\,(R^{d_1d_2d_3})_{\cA}{}^{\cB} - \frac{f_d{}^{c_1c_2 d}}{9-n}\,(t_0)_\cA{}^\cB
\nn\\
 &= \begin{pmatrix} -f_a{}^{c_1c_2 b} & 0 \\ -\frac{4!\,f_{[a_1a_2}{}^{[c_1}\,\delta^{c_2]b}_{a_3a_4]}}{\sqrt{4!}} & -4\,f_{[a_1}{}^{c_1c_2 [b_1}\,\delta^{b_2b_3b_4]}_{a_2a_3a_4]} \end{pmatrix} ,
\end{align}
where $(t_0)_\cA{}^\cB\equiv 2\,\delta_\cA^\cB$\,. 
The invariance of the bilinear form under $E_n\times \lR^+$ transformations leads to the following identity:
\begin{align}
 \langle T_C\circ T_A,\,T_B\rangle_\cD + \langle T_A,\,T_C\circ T_B\rangle_\cD + (T_C)_\cD{}^{\cE}\,\langle T_A,\, T_B\rangle_\cE =0\,.
\label{eq:U-inv}
\end{align}

\paragraph{Lie algebra of the Drinfel'd double:}
If we decompose the generators as $\{T_a\}=\{T_{\dot{a}},T_z\}$ and $\{T^{ab}\}=\{T^{\dot{a}\dot{b}},\,T^{\dot{a} z}\}$ ($\dot{a}=1,\dotsc,n-1$) and require
\begin{align}
 f_{ab}{}^{z} = 0\,,\quad f_{az}{}^{b} = 0 \,,\quad f_z{}^{b_1b_2b_3} = 0 \,,\quad f_{\dot{a}}{}^{\dot{b}_1\dot{b}_2\dot{b}_3}=0\,,
\label{eq:reduction-ansatz}
\end{align}
the subalgebra spanned by
\begin{align}
 (T_{\dot{A}})\equiv (T_{\dot{a}},\,T^{\dot{a}}) \qquad (T^{\dot{a}}\equiv T^{\dot{a}z})
\label{eq:Drinfeld-embedding1}
\end{align}
becomes
\begin{align}
\begin{split}
 T_{\dot{a}}\circ T_{\dot{b}} = f_{\dot{a}\dot{b}}{}^{\dot{c}}\,T_{\dot{c}} \,,\quad
 T_{\dot{a}}\circ T^{\dot{b}} = \tilde{f}^{\dot{b}\dot{c}}{}_{\dot{a}}\,T_{\dot{c}} - f_{\dot{a}\dot{c}}{}^{\dot{b}}\, T^{\dot{c}} = - T^{\dot{b}}\circ T_{\dot{a}}\,, \quad
 T^{\dot{a}}\circ T^{\dot{b}} = \tilde{f}^{\dot{a}\dot{b}}{}_{\dot{d}}\, T^{\dot{d}}\,,
\end{split}
\label{eq:D-reduction}
\end{align}
where $\tilde{f}^{\dot{b}\dot{c}}{}_{\dot{a}}\equiv -f_{\dot{a}}{}^{\dot{b}\dot{c}z}$\,. 
This is precisely the Lie algebra of the Drinfel'd double. 
Moreover, we can easily see that the bilinear form reduces to that of the Drinfel'd double,
\begin{align}
 \langle T_{\dot{a}},\, T^{\dot{b}}\rangle \equiv \langle T_{\dot{a}},\, T^{\dot{b}}\rangle_z = \delta^{\dot{b}}_{\dot{a}}\,. 
\end{align}
The invariance \eqref{eq:U-inv} reduces to the standard $ad$-invariance,
\begin{align}
 \langle T_{\dot{C}}\circ T_{\dot{A}},\,T_{\dot{B}}\rangle + \langle T_{\dot{A}},\,T_{\dot{C}}\circ T_{\dot{B}}\rangle = 0\,.
\end{align}
In this sense, the Leibniz algebra $\cE_n$ is an extension of the Lie algebra of the Drinfel'd double. 

It is noted that there exist certain Drinfel'd doubles which are not straightforwardly embedded into the $\cE_n$ algebra. 
When the assumption \eqref{eq:reduction-ansatz} is satisfied, the Leibniz identity \eqref{eq:Leibniz-3} for the restricted generators $T_{\dot{A}}$ is automatically satisfied. 
However, if we require the Leibniz identity \eqref{eq:Leibniz-3} for the full $\cE_n$ generators, \eqref{eq:Leibniz-3} is equivalent to $f_{d_1d_2}{}^a\, f_c{}^{d_1d_2 b}=0$\,.
Then, even under the assumption \eqref{eq:reduction-ansatz}, we obtain a constraint
\begin{align}
 f_{\dot{c}_1\dot{c}_2}{}^{\dot{a}}\, \tilde{f}^{\dot{c}_1\dot{c}_2}{}_{\dot{b}} = 0\,,
\label{eq:unimodular}
\end{align}
for the structure constants of the Drinfel'd double. 
As we discuss in section \ref{sec:discussion}, in the context of the Yang--Baxter (YB) deformation, the condition \eqref{eq:unimodular} is equivalent to the requirement that the classical $r$-matrix is unimodular. 
This means that, when the classical $r$-matrix is non-unimodular, the Lie algebra of the corresponding Drinfel'd double cannot be embedded into the $\cE_n$ algebra. 
This may be related to the fact \cite{Borsato:2016ose} that the YB deformation for a non-unimodular $r$-matrix generally produces a solution of the generalized supergravity \cite{1511.05795,1605.04884}, and the fact that the embedding of the generalized supergravity into EFT is non-trivial \cite{1612.07210} (see section \ref{sec:discussion} for further discussion). 

\paragraph{$U$-duality transformation:}
Let us consider a redefinition of the $\cE_n$ generators,
\begin{align}
 T'_A = C_A{}^B\,T_B \,,
\label{eq:NAUD}
\end{align}
where $C_A{}^B$ is an element of the $E_n$ group. 
We also redefine the bilinear-form as
\begin{align}
 \langle \cdot ,\,\cdot \rangle'_\cA = C_\cA{}^\cB\,\langle \cdot ,\,\cdot \rangle_\cB\,,
\end{align}
by acting the same $E_n$ transformation in the $R_2$-representation. 
Then, the physical subalgebra is maximally isotropic even after the redefinition,
\begin{align}
 \langle T'_a,\,T'_b \rangle'_\cC = 0\,.
\end{align}
On the other hand, the $\cE_n$ algebra is transformed as
\begin{align}
 T'_A \circ T'_B = \bigl[\Theta'^{\hat{\alpha}}_A\, (t_{\hat{\alpha}})_B{}^C + \theta'_A\,(t_0)_B{}^C\bigr]\,T'_C\,,
\end{align}
where we have defined
\begin{align}
 \Theta'^{\hat{\alpha}}_A \equiv C_A{}^B\,\Theta_B^{\hat{\beta}}\,C_{\hat{\beta}}{}^{\hat{\alpha}}\,,\quad 
 C_A{}^C\, (t_{\hat{\alpha}})_C{}^D\, (C^{-1})_D{}^B \equiv C_{\hat{\alpha}}{}^{\hat{\beta}}\,(t_{\hat{\beta}})_A{}^B \,,\quad
 \theta'_A \equiv C_A{}^B\,\theta_B\,.
\end{align}

In fact, the particular forms of $\Theta^{\hat{\alpha}}_A$ and $\theta_A$ given in \eqref{eq:En-assumption} are not preserved under a general $U$-duality transformation. 
For example, we are assuming $[\Theta_a]_{c_1c_2c_3}=0$\,, but it can appear under a general redefinition (see section \ref{sec:examples} for such an example). 
The situation is the same as the Drinfel'd double. 
In the case of the Drinfel'd double, an extension of the algebra including the non-vanishing $H$-flux (which corresponds to $[\Theta_a]_{c_1c_2c_3}$) has been discussed in \cite{1810.11446}, but here we do not consider such extension. 
Rather, we restrict the $U$-duality transformation such that $\Theta'^{\hat{\alpha}}_A$ and $\theta'_A$ have the same form as \eqref{eq:En-assumption} by using new structure constants $f'_{ab}{}^c$ and $f'_a{}^{b_1b_2b_3}$\,. 
Even under such restriction, the allowed $U$-duality symmetry is much larger than the case of the PL $T$-duality. 

\section{Generalized frame fields}
\label{sec:frame}

In this section, we present a systematic construction method of the generalized frame fields $E_A{}^I$, which is analogous to the one known in the PL $T$-duality. 
Then, by following the approach of \cite{hep-th:9710163}, we show that the $E_A{}^I$ indeed satisfy the desired relation,
\begin{align}
 \gLie_{E_A} E_B{}^I = - X_{AB}{}^C\,E_C{}^I\,,
\end{align}
where $X_{AB}{}^C$ is the structure constant $\cF_{AB}{}^C$ of the Leibniz algebra $\cE_n$\,. 

Let us prepare a set of generators $T_a$ associated with a maximal isotropic subalgebra. 
As already explained, the subalgebra is a Lie algebra, and we can parameterize an element of the Lie group $G$ as usual, e.g., $g = \Exp{x^a\,T_a}$\,. 
We define the left-/right-invariant 1-forms/vectors as
\begin{align}
 g^{-1}\,\rmd g \equiv \ell_i^a\,T_a\,\rmd x^i\,,\qquad
 \rmd g\,g^{-1} \equiv r_i^a\,T_a\,\rmd x^i\,,\qquad 
 \ell^a_i\,v^i_b = r^a_i\,e^i_b = \delta^a_b\,,
\end{align}
which satisfy
\begin{align}
 [v_a,\,v_b]^i = f_{ab}{}^c\,v_c^i\,,\qquad
 [e_a,\,e_b]^i = -f_{ab}{}^c\,e_c^i\,.
\end{align}
Then, we define the action of $g^{-1}(x)\equiv \Exp{h(x)}$ on $T_A$ as
\begin{align}
 g^{-1}(x) \circ T_A &\equiv 1 + h \circ T_A + \frac{1}{2!}\,h \circ (h \circ T_A) + \frac{1}{3!}\,h \circ (h \circ (h \circ T_A)) + \cdots 
\nn\\
 &\equiv M_A{}^B(x)\, T_B \,.
\label{eq:M-def}
\end{align}
Since the infinitesimal transformation is an $E_n\times \lR^+$ transformation of the lower-triangular form \eqref{eq:Tc-matrix}, the matrix $M_A{}^B$ can be generally parameterized as
\begin{align}
 (M_A{}^B) &= \begin{pmatrix} \delta_a^c & 0 \\ -\frac{\Pi^{a_1a_2 c}}{\sqrt{2!}} & \delta^{a_1a_2}_{c_1c_2} \end{pmatrix} \begin{pmatrix} a_c{}^b & 0 \\ 0 & (a^{-1})_{[b_1}{}^{c_1}\,(a^{-1})_{b_2]}{}^{c_2} \end{pmatrix}
\nn\\
 &=\begin{pmatrix} a_a{}^b & 0 \\ -\frac{\Pi^{a_1a_2 c}\,a_c{}^b}{\sqrt{2!}} & (a^{-1})_{[b_1}{}^{a_1}\,(a^{-1})_{b_2]}{}^{a_2} \end{pmatrix}.
\end{align}
Then, we define the generalized frame fields as
\begin{align}
 (E_A{}^I) \equiv M_A{}^B\,L_B{}^I
 = \begin{pmatrix} e_a^i & 0 \\ -\frac{\Pi^{a_1a_2 c}\,e_c^i}{\sqrt{2!}} & r_{[i_1}^{[a_1}\,r_{i_2]}^{a_2]} \end{pmatrix},
\label{eq:EAI}
\end{align}
where the matrix $L_A{}^I$ is defined by
\begin{align}
 (L_A{}^I) \equiv \begin{pmatrix} v_a^i & 0 \\ 0 & \ell_{[i_1}^{[a_1}\,\ell_{i_2]}^{a_2]} \end{pmatrix},
\end{align}
and we have used $\ell^a_i = a_b{}^a\,r^b_i$ in the second equality of \eqref{eq:EAI}. 
This matrix $E_A{}^I$ plays the role of the desired generalized frame fields, as we show below. 
For this purpose, let us find several identities by following \cite{hep-th:9710163}. 

\paragraph{Differential identities:}
Differentiating the definition \eqref{eq:M-def} of the matrix $M_A{}^B$, we obtain
\begin{align}
 \partial_i g^{-1}(x) \circ T_A = \partial_i M_A{}^B(x)\, T_B\,.
\label{eq:dg-1TA}
\end{align}
The left-hand side can be evaluated as
\begin{align}
 &\partial_i g^{-1} \circ T_A = -g^{-1}\circ \partial_i g\circ g^{-1}\circ T_A = -\bigl(\ell_i^d\,T_d\bigr)\circ \bigl(M_A{}^B\,T_B\bigr) = -\ell_i^d\,M_A{}^B\,(T_d)_B{}^C\,T_C 
\nn\\
 &= \ell_i^d \begin{pmatrix} a_a{}^b\,f_{bd}{}^c & 0 \\ -\frac{\Pi^{a_1a_2 c}\,a_c{}^b\,f_{bd}{}^c + (a^{-1})_{[b_1}^{~~a_1}\,(a^{-1})_{b_2]}^{~~a_2}\,f_d{}^{b_1b_2c}}{\sqrt{2!}} & 2\,(a^{-1})_{[b_1}^{~~a_1}\,(a^{-1})_{b_2]}^{~~a_2}\,f_{d[c_1}{}^{[b_1}\,\delta^{b_2]}_{c_2]} \end{pmatrix} T_C\,,
\end{align}
and \eqref{eq:dg-1TA} gives the following identities:
\begin{align}
 D_c a_a{}^b = a_a{}^d\,a_c{}^e \,f_{de}{}^b \,,\qquad 
 D_c \Pi^{a_1a_2a_3} = (a^{-1})_{b_1}{}^{a_1}\,(a^{-1})_{b_2}{}^{a_2}\,(a^{-1})_{b_3}{}^{a_3}\,a_c{}^d\,f_d{}^{b_1b_2b_3} \,.
\end{align}

\paragraph{Algebraic identities:}
In order to find further relations, we consider the identity
\begin{align}
 (g\circ T_A) \circ (g\circ T_B) = g\circ (T_A \circ T_B)\,,
\label{eq:gTAgTB-TC}
\end{align}
which follows from the Leibniz identity. 
For convenience, we decompose this identity as
\begin{align}
 \langle (g\circ T_A) \circ (g\circ T_B) , \, T_C \rangle_\cD
 = \langle g\circ (T_A \circ T_B) , \, T_C \rangle_\cD \,. 
\end{align}
The component $\{{}_A,\,{}_B,\,{}_C,\,{}_\cD\}=\{{}^{a_1a_2},\,{}_b,\,{}_c,\,{}_{d}\}$ or $\{{}_A,\,{}_B,\,{}_C,\,{}_\cD\}=\{{}_a,\,{}^{b_1b_2},\,{}_c,\,{}_{d}\}$ leads to
\begin{align}
 (a^{-1})_a{}^e\,(a^{-1})_b{}^f\,a_g{}^c\,f_{ef}{}^g = f_{ab}{}^c\,. 
\end{align}
On the other hand, the component $\{{}_A,\,{}_B,\,{}_C,\,{}_\cD\}=\{{}_a,\,{}^{b_1b_2},\,{}^{c_1c_2},\,{}_{d}\}$ additionally requires
\begin{align}
 a_a{}^e\, (a^{-1})_{f_1}{}^{b_1}\,(a^{-1})_{f_2}{}^{b_2}\,(a^{-1})_{f_3}{}^{b_3}\, f_e{}^{f_1f_2f_3} = f_a{}^{b_1b_2b_3} + 3\, f_{ac}{}^{[b_1}\,\Pi^{b_2b_3]c} \,,
\end{align}
and the component $\{{}_A,\,{}_B,\,{}_C,\,{}_\cD\}=\{{}^{a_1a_2},\,{}_b,\,{}^{c_1c_2}\,{}_{d}\}$ also requires
\begin{align}
 f_{e_1e_2}{}^{[a_1}\,\delta^{a_2]}_{b}\,\delta^{[c_1}_d\,\Pi^{c_2]e_1e_2} =0\quad\Leftrightarrow\quad f_{ab}{}^c\,\Pi^{abd} =0 \,. 
\label{alg-F}
\end{align}
The component $\{{}_A,\,{}_B,\,{}_C,\,{}_\cD\}=\{{}^{a_1a_2},\,{}^{b_1b_2},\,{}_{c},\,{}_{d}\}$ further gives\footnote{For $n\leq 4$, \eqref{eq:alg-Q} automatically follows from \eqref{alg-F}. Indeed, for $n\leq 4$, \eqref{eq:alg-Q} is equivalent to a trivial identity $f_{e_1e_2}{}^{a_1}\,\Pi^{[a_2b_1b_2}\,\delta^{e_1e_2]}_{cd} - f_{e_1e_2}{}^{a_2}\,\Pi^{[a_1b_1b_2}\,\delta^{e_1e_2]}_{cd}=0$ under \eqref{alg-F}.}
\begin{align}
 3\,\bigl(f_{e[c}{}^{a_1}\,\delta_{d]}^{[a_2}\,\Pi^{b_1b_2]e}
 - f_{e[c}{}^{a_2}\,\delta_{d]}^{[a_1}\,\Pi^{b_1b_2]e} \bigr)
 + f_{cd}{}^{[a_1}\,\Pi^{a_2]b_1b_2} = 0 \,.
\label{eq:alg-Q}
\end{align}
Finally, the component $\{{}_A,\,{}_B,\,{}_C,\,{}_\cD\}=\{{}^{a_1a_2},\,{}^{b_1b_2},\,{}^{c_1c_2},\,{}_{d}\}$ gives
\begin{align}
 f_d{}^{b_1b_2c}\,\Pi^{a_1a_2d}
 -3\,f_d{}^{a_1a_2[b_1}\,\Pi^{b_2c]d}
 = 3\,f_{de}{}^{[c}\,\Pi^{b_1b_2]d}\,\Pi^{a_1a_2e}
 - 4\,f_{de}{}^{[a_1}\,\Pi^{a_2]d[b_1}\,\Pi^{b_2]ec} \,,
\label{eq:alg-R}
\end{align}
and they are all identities coming from \eqref{eq:gTAgTB-TC}. 

\paragraph{Computation of $X_{AB}{}^C$:}
By using the differential and algebraic identities, we can easily show
$X_{a}{}^{b_1b_2}{}^c = f_a{}^{b_1b_2 c}$ and $X^{a_1a_2}{}_b{}^c = -f_{b}{}^{a_1a_2}{}^c$\,. 
The derivation of
\begin{align}
 X^{a_1a_2b_1b_2c} = 0\,,\qquad X^{a_1a_2b_1b_2}{}_{c_1c_2} = -4\,f_{d}{}^{a_1a_2}{}^{[b_1}\, \delta^{b_2]d}_{c_1c_2}\,,
\end{align}
requires a slightly longer computation. 
The former requires the identity \eqref{eq:alg-R} while the latter requires \eqref{eq:alg-Q}. 
In this way, we have shown the desired relation $X_{AB}{}^C=\cF_{AB}{}^C$\,. 

In summary, by using the $\cE_n$ algebra, we have explained a systematic construction of the generalized frame fields $E_A{}^I$, which satisfy the algebra of $\cE_n$ by means of the generalized Lie derivative. 
The construction is a straightforward extension of the procedure known in the PL $T$-duality, and we expect that this extension plays an important role in formulating the $U$-duality extension of the PL $T$-duality. 

\section{Examples of $\cE_n$ algebra}
\label{sec:examples}

\subsection{3D algebra $\cE_2$}

When $n=2$, we obtain a three-dimensional algebra with generators $\{T_A\} = \{T_1,\,T_2,\,T^{12}\}$\,. 
By denoting $f_{12}{}^1=a$ and $f_{12}{}^2=b$, we obtain
\begin{align}
 T_1\circ T_2 &= a\,T_1 + b\,T_2 = [T_1,\,T_2] \,,
\\
 T_1\circ T^{12} &= - b\, T^{12} \,, \quad
 T_2\circ T^{12} = a\, T^{12} \,,\quad
 T^{12}\circ T_A = 0\,.
\end{align}
The non-vanishing components of the bilinear form are
\begin{align}
 \langle T_1,\, T^{12}\rangle_2 = -1\,,\qquad 
 \langle T_2,\, T^{12}\rangle_1 = 1 \,. 
\label{eq:<TaTab-E2>}
\end{align}

This is not an interesting example, but it is a good example to clearly see the existence of another maximal isotropic subalgebra. 
As we can clearly see from \eqref{eq:<TaTab-E2>}, the generator $T^{12}$ has non-vanishing inner products with other generators. 
This shows that the Abelian algebra generated by $\{\sfT_\sfa\}=\{T^{12}\}$ is another maximal isotropic subalgebra. 
Similarly, $\cE_n$ always has two types of maximal isotropic subalgebras with dimension $n$ and $n-1$\,. 

\subsection{6D algebra $\cE_3$}

The algebra $\cE_3$ is a six-dimensional algebra with generators $\{T_A\}=\{T_1,\,T_2,\,T_3,\,T^{12},\,T^{13},\,T^{23}\}$\,. 
The structure constants $f_{ab}{}^c$ have 9 components and $f_a{}^{bcd}$ have 3 components. 
According to the Bianchi classification, the 3D Lie algebra $f_{ab}{}^c$ has been classified. 
It is interesting to classify the additional structure constants $f_a{}^{bcd}$ for each physical 3D Lie algebra. 

\subsection{10D algebra $\cE_4$}

\paragraph{M-theory frame I:}
In $n=4$, the algebra $\cE_4$ is ten-dimensional and the structure is much richer. 
As a particular example, we here consider the case
\begin{align}
 f_{ab}{}^c=0\,,\quad f_1{}^{234}=a\,,\quad f_1{}^{134}=b\,,\quad f_2{}^{234}=c\,,\quad f_2{}^{134}=d\,,
\label{eq:example1}
\end{align}
which satisfies the Leibniz identity. 
If we introduce an additional non-vanishing component for $f_a{}^{b_1b_2b_3}$, the Leibniz identity is broken, and in that sense it contains a maximal set of components under $f_{ab}{}^c=0$\,. 
Using the generators $T_a$\,, we parameterize an element of the physical subgroup as $g= \Exp{x^a\,T_a}$. 
As it is Abelian, the left-/right-invariant forms are trivial,
\begin{align}
 \ell=r= T_a\,\rmd x^a \,. 
\end{align}
On the other hand, the tensor $\Pi^{i_1i_2i_3}\equiv e^{i_1}_{a_1}\,e^{i_2}_{a_2}\,e^{i_3}_{a_3}\,\Pi^{a_1a_2a_3}$ has the form
\begin{align}
 \Pi = \bigl[(b\, x^1+d\,x^2)\,\partial_1 + (a\, x^1 + c\,x^2)\,\partial_2\bigr]\wedge\partial_3\wedge\partial_4\,.
\end{align}
By construction, the $R$-flux $X^{a_1a_2b_1b_2c}$ should vanish, and it satisfies
\begin{align}
 \Pi^{i_1i_2k}\, \partial_k \Pi^{j_1j_2l} - 3\,\Pi^{[j_1j_2|k}\, \partial_k \Pi^{i_1i_2|l]} = 0 \,.
\end{align}
In order for this to be a Nambu--Poisson tensor, the algebraic or quadratic identity
\begin{align}
 \Pi^{k[i_1i_2}\, \Pi^{i_3] j l} + \Pi^{l[i_1i_2}\, \Pi^{i_3] jk} =0
\end{align}
should be satisfied \cite{hep-th:9301111} (see also \cite{math:9901047}). 
In this example, it is indeed satisfied and the above $\Pi$ is a Nambu--Poisson tensor. 
In general, we have not checked the quadratic identity, but it may follow from a certain requirement such as the Leibniz identity. 

By using the trivial right-invariant vector and the Nambu--Poisson structure, the generalized frame fields become
\begin{align}
 E_A{}^I =\begin{pmatrix} e_a^i & 0 \\ -\tfrac{\Pi^{a_1a_2 b}\,e_b^i}{\sqrt{2!}} & r^{[a_1}_{[i_1}\,r^{a_2]}_{i_2]} \end{pmatrix} 
 =\begin{pmatrix} \delta_a^i & 0 \\ -\tfrac{\Pi^{a_1a_2 i}}{\sqrt{2!}} & \delta^{a_1a_1}_{i_1i_2} \end{pmatrix} .
\label{eq:M-twist}
\end{align}
As we have generally proven, this satisfies the relation $\gLie_{E_A}E_B{}^I = - \cF_{AB}{}^C\,E_C{}^I$ for the structure constants given in \eqref{eq:example1}. 

\paragraph{M-theory frame II:}
Let us consider a redefinition,
\begin{align}
 \footnotesize{\begin{pmatrix} T'_1 \\ T'_2 \\ T'_3 \\ T'_4 \\ T'^{12}\\ T'^{13}\\ T'^{14}\\ T'^{23}\\ T'^{24}\\ T'^{34} \end{pmatrix}
 \equiv \begin{pmatrix}
 1 & 0 & 0 & 0 & 0 & 0 & 0 & 0 & 0 & 0 \\
 0 & 0 & 0 & 0 & 0 & 0 & 0 & 0 & 1 & 0 \\
 0 & 0 & 0 & 0 & 0 & 0 & 0 & 0 & 0 & 1 \\
 0 & 0 & 0 & 0 & 0 & 0 & 0 & -1 & 0 & 0 \\
 0 & 0 & 0 & 0 & 0 & 1 & 0 & 0 & 0 & 0 \\
 0 & 0 & 0 & 0 & -1 & 0 & 0 & 0 & 0 & 0 \\
 0 & 0 & 0 & 0 & 0 & 0 & 1 & 0 & 0 & 0 \\
 0 & 0 & 0 & 1 & 0 & 0 & 0 & 0 & 0 & 0 \\
 0 & 1 & 0 & 0 & 0 & 0 & 0 & 0 & 0 & 0 \\
 0 & 0 & 1 & 0 & 0 & 0 & 0 & 0 & 0 & 0 
\end{pmatrix}
 \begin{pmatrix} T_1 \\ T_2 \\ T_3 \\ T_4 \\ T^{12}\\ T^{13}\\ T^{14}\\ T^{23}\\ T^{24}\\ T^{34} \end{pmatrix}} . 
\end{align}
This is a map
\begin{align}
 \text{M-theory} \overset{{\text{reduction} \atop \text{on $x^4$}}\vphantom{\Big|}}{\rightarrow} \text{Type IIA} 
 \overset{{\text{$T$-dualities} \atop \text{along $x^3$ and $x^2$}}\vphantom{\Big|}}{\rightarrow} \text{Type IIA} 
 \overset{{\text{recovery} \atop \text{of $x^4$}}\vphantom{\Big|}}{\rightarrow} \text{M-theory} \,,
\end{align}
corresponding to a double Abelian $T$-duality, and it is a particular $U$-duality transformation. 

After the redefinition, we find the new physical generators satisfy
\begin{align}
 [T'_a,\,T'_b] = f'_{ab}{}^c\,T'_c + \frac{1}{2!}\,f'_{abc_1c_2}\,T'^{c_1c_2}\,,
\end{align}
where $f'_{a_1\cdots a_4}\equiv f'_{[a_1\cdots a_4]}$ and
\begin{align}
 f'_{13}{}^1 = b\,,\quad f'_{23}{}^2 = -c\,,\quad f'_{34}{}^4 = c\,,\quad f'_3{}^{124} = -d\,,\quad
 f'_{1234} = - a\,.
\end{align}
The component $f'_{abc_1c_2}$ is not allowed in the $\cE_4$ algebra, and we can consider this $U$-duality transformation only when $a=0$\,.\footnote{The flux $f_1{}^{234}=a$ corresponds to the $Q$-flux $Q_1{}^{23}=-a$ in type IIA theory and the double $T$-duality transforms it to the $H$-flux $H_{123}=-a$\,. The 11D uplift corresponds to $f'_{1234}=-a$\,.} 
Moreover, the algebra of other generators further requires $c=0$\,.\footnote{The reason may be understood as follows. Originally, the $\Pi^{a_1a_2a_3}$ has the $x^2$-dependence, but under the double $T$-dualities, $x^2$ is mapped to $y_{24}$\,. The dependence on the dual coordinate breaks our assumption \eqref{eq:M-section}. Accordingly, the resulting algebra has a different form from $\cE_4$\,.} 
Under $a=c=0$, the $U$-duality converts the structure constants of the $\cE_4$ algebra as
\begin{align}
 f'_{13}{}^1 = b\,,\qquad f'_3{}^{124} = -d\,.
\end{align}
Again, we can easily construct the generalized frame fields realizing this algebra. 

\paragraph{Type IIB frame:}
Let us consider another redefinition of the $\cE_4$ generators,
\begin{align}
 \{\sfT_\sfA\} \equiv \{ T_1\,,\, T_2\,,\, T^{34}\,,\, T^{14}\,,\, T^{24}\,,\, T_3\,,\, -T^{13}\,,\, -T^{23}\,,\, T_4\,,\, T^{12}\}\,.
\label{eq:B-generators}
\end{align}
This map has been considered in \cite{1701.07819}, which connects the M-theory picture and the type IIB picture (see \cite{hep-th:0402140,1311.5109} for earlier discussion). 
This is not a $U$-duality transformation but rather corresponds to a change in the picture, from M-theory to type IIB theory.\footnote{One can see that the bilinear form is not invariant under the transformation (see \cite{1701.07819} for the transformation rule of the index $\cA$ under this redefinition).}

In type IIB theory, we can decompose the $R_1$-representation (for $n\leq 5$) as
\begin{align}
 \{\sfT_\sfA\} = \bigl\{ \sfT_\sfa,\,\sfT^{\sfa}_\alpha,\,\tfrac{\sfT^{\sfa_1\sfa_2\sfa_3}}{\sqrt{3!}}\bigr\}\qquad 
 \bigl(\sfa=1,\dotsc,n-1,\quad \alpha=1,2,\quad \sfT^{\sfa_1\sfa_2\sfa_3}=\sfT^{[\sfa_1\sfa_2\sfa_3]}\bigr)\,,
\end{align}
and we understand the above redefinition as
\begin{align}
\begin{split}
 \{\sfT_\sfa\} &= \{T_1\,,\, T_2\,,\, T^{34}\}\,,\qquad 
 \{\sfT_1^\sfa\} = \{T^{14}\,,\, T^{24}\,,\, T_3\}\,,
\\
 \{\sfT_2^\sfa\} &= \{-T^{13}\,,\, -T^{23}\,,\, T_4\}\,,\qquad 
 \sfT^{123} = T^{12}\,.
\end{split}
\end{align}
Then, we can see that the set of generators $\{\sfT_\sfa\}$ forms a maximal isotropic subalgebra,
\begin{align}
 [\sfT_\sfa,\,\sfT_\sfb] = \sff_{\sfa\sfb}{}^\sfc\,\sfT_\sfc \qquad 
 \bigl(\sff_{13}{}^1 = b\,,\quad \sff_{13}{}^2 = a\,,\quad \sff_{23}{}^1 = d\,,\quad \sff_{23}{}^2 = c \bigr)\,. 
\end{align}
This algebra should be regarded as the physical subalgebra in type IIB theory. 
Although the entire algebra in the type IIB picture has not been established, it seems that this example does not contain any dual structure constants, which may have the form $\sff_\sfa{}^{(\alpha\beta)}$ or $\sff_{\sfa}^{\vphantom{\sfb}}{}^{[\sfb_1\sfb_2]}_\beta$\,. 

If we restrict ourselves to the case $a=0$, $b=-1$, $c=1$, and $d=0$\,, the algebra is Bianchi type $\bm{6_0}$,
\begin{align}
 [\sfT_3,\,\sfT_1] = \sfT_1\,,\qquad [\sfT_3,\,\sfT_2] = -\sfT_2\,.
\end{align}
In this case, by using a supergravity solution obtained in \cite{hep-th:0205245} (which has the symmetry of the Bianchi type $\bm{6_0}$), we can perform a $U$-duality extension of the PL $T$-duality. 
Namely, in the M-theory picture, we can construct a solution of EFT that is twisted by the matrix \eqref{eq:M-twist} with $a=0$, $b=-1$, $c=1$, and $d=0$\,. 
Under the change of the generators, the solution is mapped to the type IIB solution of \cite{hep-th:0205245}. 
However, this is not so interesting because it is nothing more than the straightforward 11D uplift of the PL $T$-duality. 
In the type IIA picture, \eqref{eq:example1} reduces to the Lie algebra of the Drinfel'd double [recall \eqref{eq:D-reduction} and choose $z=4$]
\begin{align}
 f_{ab}{}^c=0\,, \quad \tilde{f}^{13}{}_1= 1\,,\quad \tilde{f}^{23}{}_2=-1\,, 
\end{align}
where the physical algebra is Abelian and the dual algebra is Bianchi type $\bm{6_0}$\,. 
Then, the redefinition \eqref{eq:B-generators} corresponds to a non-Abelian $T$-duality. 
In order to find genuinely $U$-duality examples, it is important to study the detailed classification of the $\cE_n$ algebra. 

\section{Summary and Discussion}
\label{sec:discussion}

\paragraph{Summary:}
When we perform the PL $T$-duality, a systematic construction of the generalized frame fields satisfying $\gLie_{E_A}E_B{}^M = -\cF_{AB}{}^C\,E_C{}^M$ with a constant $\cF_{AB}{}^C$ is useful. 
In this paper, by considering the $U$-duality extension of the PL $T$-duality, we have proposed a Leibniz algebra $\cE_n$\,, which extends the Lie algebra of the Drinfel'd double. 
Then, we have shown that this provides a systematic way to construct the generalized frame fields in EFT, which satisfy the $\cE_n$ algebra $\gLie_{E_A}E_B{}^I= -\cF_{AB}{}^C\,E_C{}^I$ by means of the generalized Lie derivative in EFT. 

\paragraph{Straightforward extensions:}

In this paper, we have concentrated on the case $n\leq 4$\,, but the extension to higher $n$ will be straightforward. 
Since the generators $T_A$ are transforming in the $R_1$-representation, for higher $n$, we introduce the following generators:
\begin{align}
 \{T_A\} = \{T_a,\,T^{a_1a_2},\,T^{a_1\cdots a_5},\,T^{a_1\cdots a_7, a},\,\cdots \}\,. 
\end{align}
According to the success of the $E_{11}$ conjecture \cite{hep-th:0104081,hep-th:0307098}, it will be possible to extend $n$ up to $n=11$\,. 
Here we have almost restricted ourselves to the M-theory picture, but if we consider the type IIB picture, the generators are parameterized as
\begin{align}
 \{\sfT_\sfA\} = \{\sfT_\sfa,\,\sfT^{\sfa}_\alpha,\,\sfT^{\sfa_1\sfa_2\sfa_3},\,\sfT^{\sfa_1\cdots \sfa_5}_\alpha,\, \sfT^{\sfa_1\cdots \sfa_6, \sfa},\,\cdots \}\,. 
\end{align}
The invariant bilinear forms in both the M-theory/type IIB pictures are also well studied in EFT (see \cite{1708.06342} for $n\leq 7$). 
The algebra should always have the form
\begin{align}
 T_A \circ T_B = \bigl[\Theta_A^{\hat{\alpha}}\, (t_{\hat{\alpha}})_B{}^C + \theta_A\,(t_0)_B{}^C\bigr]\,T_C\,,
\end{align}
and what we need to do for higher $n$ will be to consistently find the constants $\Theta_A^{\hat{\alpha}}$ and $\theta_A$\,. 
The construction method of $E_A{}^I$ will also be straightforward to extend to higher $n$\,. 

\paragraph{Towards non-Abelian $U$-duality:}
The most interesting application of our result is the $U$-duality extensions of the PL $T$-duality, which may be called the Nambu--Lie $U$-duality. 
In order to find non-trivial examples of such $U$-duality, the decompositions of the $\cE_n$ algebra into the physical and the dual subalgebras need to be classified. 
In the case of the Drinfel'd double, such decomposition is known as the Manin triple, and the classification for the six-dimensional case has been worked out in \cite{math:0202210}. 
The extension of such classification for each $\cE_n$ algebra is important. 
A major difference from the case of the Drinfel'd double is in the existence of the two types of subalgebras with dimensions $n$ and $n-1$\,. 
Another difference is that the dual algebra of $\cE_n$ (generated by $T^{a_1a_2}$) is not maximally isotropic and accordingly is a Leibniz algebra. 
Namely, unlike the case of the Drinfel'd double, the $\cE_n$ algebra is decomposed into an $n$-dimensional physical Lie algebra and a $(D_n-n)$-dimensional dual Leibniz algebra. 
It is also noted that the $\cE_n$ algebra in the M-theory picture and that in the type IIB picture may not be exactly the same in general. 
In the M-theory picture, we introduced the structure constants $f_{ca}{}^b$ and $f_c{}^{a_1a_2a_3}$ corresponding to the $E_n$ generators $K^a{}_b$ and $R_{a_1a_2a_3}$ but do not introduce $f_c{}_{a_1a_2a_3}$, which corresponds to $R^{a_1a_2a_3}$\,. 
On the other hand, in the type IIB picture, we may introduce $\sff_{\sfc\sfa}{}^{\sfb}$, $\sff_\sfc{}^{(\alpha\beta)}$, and $\sff_{\sfc}{}^{[\sfa_1\sfa_2]}_\alpha$ corresponding to the $E_n$ generators $\sfK^\sfa{}_\sfb$, $\sfR_{(\alpha\beta)}$, and $\sfR^\alpha_{\sfa_1\sfa_2}$, but will not introduce $\sff_{\sfc\vphantom{\sfb}}^{\vphantom{\alpha}}{}_{[\sfa_1\sfa_2]}^\alpha$, which is associated with $\sfR_\alpha^{\sfa_1\sfa_2}$\,. 
Then, the number of the structure constants does not match between the two pictures. 
It may coincide after imposing the Leibniz identity, but it is not obvious and it is important to study the correspondence in detail. 

It is also important to study the flux-formulation of EFT. 
In the case of gauged DFT \cite{1109.0290,1109.4280,1201.2924,1304.1472,1705.08181,1706.08883}, the action and equations of motion are expressed purely by using the generalized flux $\cF_{ABC}$ (and additional flux $\cF_A$). 
Moreover, when the flux is constant, the equations of motion reduce to the algebraic equations \eqref{eq:DFT-eom}. 
A similar analysis has been done in \cite{1302.5419}, and the action of EFT is expressed by the generalized flux $X_{AB}{}^C$\,. 
If the equations of motion are also expressed by using the fluxes, and if they reduce to simple algebraic equations when $X_{AB}{}^C$ is constant, we can clearly see the symmetry of the non-Abelian $U$-duality. 

\paragraph{Duality in the membrane sigma model:}
It is important to study the duality symmetry also in the context of the membrane sigma model. 
Originally, the PL $T$-dualizability condition has been found in the form,
\begin{align}
 \Lie_{v_a}E_{mn}= -\tilde{f}^{bc}{}_a\,E_{mp}\,v_b^p\, v_c^q\,E_{qn} \,,
\label{eq:Lie-E}
\end{align}
where $E_{mn}\equiv g_{mn}-B_{mn}$\,. 
By solving the differential equation with the help of the Drinfel'd double, the twist matrix \eqref{eq:PL-dualizable-BG} has been obtained. 
The condition \eqref{eq:Lie-E} shows that the equations of motion of the string sigma model are expressed as a Maurer--Cartan equation,
\begin{align}
 \rmd J_a - \frac{1}{2}\,\tilde{f}_a{}^{bc}\, J_b\wedge J_c = 0\,,\qquad 
 J_a \equiv v_a^m\,\bigl(g_{mn}\, *\rmd x^n - B_{mn}\, \rmd x^n\bigr) \,,
\label{eq:MC-string}
\end{align}
and this plays an important role in realizing the PL $T$-duality as a symmetry in the equation of motion of string theory. 
When the matrix $E_{mn}$ is invertible, \eqref{eq:Lie-E} is equivalent to
\begin{align}
 \Lie_{v_a}E^{mn}= \tilde{f}^{bc}{}_a\,v_b^m\, v_c^n \,,
\end{align}
and if we define a dual metric $\tilde{g}_{mn}$ and a bi-vector $\beta^{mn}$ through the relation
\begin{align}
 \tilde{g}^{mn}+\beta^{mn} = \bigl[(g+B)^{-1}\bigr]^{mn}\,,
\end{align}
the dualizability condition is expressed as
\begin{align}
 \Lie_{v_a} \tilde{g}_{mn}=0\,,\qquad \Lie_{v_a}\beta^{mn}=-\tilde{f}^{bc}{}_a\,v_b^m\, v_c^n \,. 
\label{eq:Lie-g-beta}
\end{align}
In fact, we can easily find a similar relation in our setup. 
If we define the generalized metric as $\cM_{IJ} = \abs{r}^{\frac{2}{9-n}}\, E_I{}^A\,E_J{}^B\,\hat{\cM}_{AB}$ by using the twist matrix \eqref{eq:EAI} and a diagonal constant metric $\hat{\cM}_{IJ}$ made of an invariant metric $\kappa_{ab}$ of the physical subgroup $G$, we find that the dual metric $\tilde{g}_{ij}$ (i.e.~the M-theory uplift of $\tilde{g}_{mn}$) and the $\Omega$-field (i.e.~the M-theory uplift of the $\beta$-field, $\Omega^{mnz}=\beta^{mn}$) are given by
\begin{align}
 \tilde{g}_{ij} = r_i^a\,r_j^b\,\kappa_{ab}\,,\qquad 
 \Omega^{i_1i_2i_3} = \Pi^{i_1i_2i_3} \equiv e_{a_1}^{i_1}\,e_{a_2}^{i_2}\,e_{a_3}^{i_3}\,\Pi^{a_1a_2a_3}\,. 
\end{align}
Then, we can show the following relation, which is the M-theory uplift of \eqref{eq:Lie-g-beta}:
\begin{align}
 \Lie_{v_a} \tilde{g}_{ij}=0\,,\qquad 
 \Lie_{v_a} \Omega^{i_1i_2i_3} = \Lie_{v_a} \Pi^{i_1i_2i_3} = f_a{}^{b_1b_2b_3}\,v_{b_1}^{i_1}\,v_{b_2}^{i_2}\,v_{b_3}^{i_3}\,.
\label{eq:U-condition}
\end{align}
It is interesting to study the implication of these relations in the context of the membrane sigma model. 
Perhaps the equations of motion of the membrane sigma model can be expressed in a similar form as \eqref{eq:MC-string}, and this may help us to discuss the non-Abelian $U$-duality in the context of the membrane sigma model. 

\paragraph{Generalized Yang--Baxter deformation:}
Another related direction is a generalization of the YB deformation \cite{hep-th:0210095,0802.3518,1308.3581,1309.5850,1401.4855}. 
As it has been observed in \cite{1705.07116}, the YB deformation is a coordinate-dependent $\beta$-deformation $\beta^{mn}\to \beta'^{mn} =\beta^{mn} + r^{mn}$, associated with a bi-Killing vector $r^{mn} \equiv r^{ab}\,v_a^m\,v_b^n$, where $r^{ab}=-r^{ba}$ is a constant matrix. 
Here, the set of vector fields $v_a^m$ satisfies the algebra $[v_a,\,v_b]^m = f_{ab}{}^c\,v_c^m$ and the Killing equation $\Lie_{v_a}(g+B)_{mn}=0$ in the undeformed background. 
In this case, the YB-deformed background satisfies the PL $T$-dualizability condition \eqref{eq:Lie-g-beta} with the dual structure constant given by
\begin{align}
 \tilde{f}^{b_1b_2}{}_a = 2\,r^{c[b_1}\,f_{ac}{}^{b_2]}\,.
\label{eq:f-tilde-YB}
\end{align}
Interestingly, when the matrix $r^{ab}$ satisfies the homogeneous classical YB equations (CYBE),
\begin{align}
 f_{d_1d_2}{}^a\,r^{bd_1}\,r^{cd_2} 
 + f_{d_1d_2}{}^b\,r^{cd_1}\,r^{ad_2}
 + f_{d_1d_2}{}^c\,r^{ad_1}\,r^{bd_2} = 0\,,
\label{eq:CYBE}
\end{align}
the YB deformation always maps a DFT solution to another DFT solution. 
The reason can be clearly understood by noticing that the YB deformation is a particular PL $T$-duality. 

Before the YB deformation, the background satisfies $\Lie_{v_a}(g+B)_{mn}=0$ and this shows that $\tilde{f}^{ab}{}_c=0$\,. 
Namely, in the original background, which is described by the generalized metric $\cH_{MN}=E_M{}^A\,E_N{}^B\,\hat{\cH}_{AB}$, the fields $E_A{}^M$, $\tilde{g}_{mn}$, and $\beta^{mn}$ have the following form:
\begin{align}
 (E_A{}^M) = \begin{pmatrix} e_a^m & 0 \\ 0 & r^a_m \end{pmatrix},\quad \tilde{g}_{mn} = r_m^a\,r_n^b\,\kappa_{ab} = \ell_m^a\,\ell_n^b\,\kappa_{ab}\,,\quad \beta^{mn}=e^m_a\,e^n_b\,\beta^{ab}\,,
\end{align}
where $\kappa_{ab}$ and $\beta^{ab}$ are constant, and $\kappa_{ab}$ is supposed to be an invariant metric of the isometry algebra. 
The YB deformation corresponds to the PL $T$-duality \eqref{eq:PL-transf} with
\begin{align}
 (C_A{}^B) = \begin{pmatrix} \delta_a^b & 0 \\ -r^{ab} & \delta^a_b \end{pmatrix}.
\label{eq:CAB-YB}
\end{align}
Under this transformation, the generators become $T'_a=T_a$ and $T'^a = T^a - r^{ab}\,T_b$ and the constant fields are transformed as $\kappa_{ab}\to \kappa'_{ab}=\kappa_{ab}$ and $\beta^{ab}\to \beta'^{ab}=\beta^{ab}+r^{ab}$\,. 
By requiring that the new generators $T'_A$ satisfy the Lie algebra of the Drinfel'd double with $\tilde{f}^{b_1b_2}{}_a$ given by \eqref{eq:f-tilde-YB}, the matrix $r^{ab}$ must be a classical $r$-matrix satisfying \eqref{eq:CYBE}. 
Then, using the systematic construction of $E'_A{}^M$, we can in principle compute the generalized frame fields
\begin{align}
 (E'_A{}^M) = \begin{pmatrix} \delta_a^b & 0 \\ \Pi^{ab}(x) & \delta^a_m \end{pmatrix}\begin{pmatrix} e_b^m & 0 \\ 0 & r^b_m \end{pmatrix},
\end{align}
and the $\beta$-field in the deformed background can be computed as
\begin{align}
 \beta'^{mn} = \beta^{mn} + \pi^{mn}\,,\qquad \pi^{mn} \equiv e_a^m\,e_b^n\,\bigl(\Pi^{ab}+r^{ab}\bigr)\,.
\end{align}
The great benefit of the YB deformation is that we do not need to compute $\pi^{mn}$. 
It is simply given by $\pi^{mn}=r^{mn}$ because $r^{mn}$ solves the differential equation $\Lie_{v_a}r^{mn}=-\tilde{f}^{bc}{}_a\,v_b^m\, v_c^n$\,, and $\pi^{mn}=r^{mn}$ is trivially satisfied at the identity $g=1$ [recall \eqref{eq:Pi=0}]. 
Thus, once we find a classical $r$-matrix, we can easily generate a new solution. 
In this sense, the YB deformation is a systematic way to perform the PL $T$-duality \eqref{eq:CAB-YB} and the homogeneous CYBE ensure that the structure of the Drinfel'd double is preserved under the deformation. 

Recently, an 11D extension of this YB deformation has been studied in \cite{1906.09052}. 
There, the YB deformation is generalized to the $\Omega$-deformation $\Omega^{i_1i_2i_3} \to \Omega'^{i_1i_2i_3} = \Omega^{i_1i_2i_3} + \rho^{i_1i_2i_3}$ associated with a tri-Killing vector
\begin{align}
 \rho^{i_1i_2i_3} = \rho^{a_1a_2a_3}\,v_{a_1}^{i_1}\,v_{a_2}^{i_2}\,v_{a_3}^{i_3}\,,
\label{eq:Pi-tri}
\end{align}
where $\rho^{a_1a_2a_3}=\rho^{[a_1a_2a_3]}$ is a certain constant. 
By assuming the Killing equations ($\Lie_{v_a}g_{ij}=0$ and $\Lie_{v_a}C_{i_1i_2i_3}=0$) in the undeformed background, the $\Omega$-deformed background satisfies the relation \eqref{eq:U-condition} with the dual structure constant given by
\begin{align}
 f_a{}^{b_1b_2b_3} = 3\,\rho^{c[b_1b_2}\,f_{ac}{}^{b_3]} \,. 
\end{align}
Similar to the case of the YB deformation, this also can be understood as a particular non-Abelian $U$-duality transformation \eqref{eq:NAUD} with
\begin{align}
 (C_A{}^B) = \begin{pmatrix} \delta_a^b & 0 \\ \tfrac{\rho^{a_1a_2b}}{\sqrt{2!}} & \delta^{a_1a_2}_{b_1b_2} \end{pmatrix}.
\end{align}
By requiring the redefined generators $T'_a=T_a$ and $T'^{a_1a_2}=T^{a_1a_2}+\rho^{a_1a_2b}\,T_b$ to satisfy the $\cE_n$ algebra, we obtain
\begin{align}
 f_{b_1b_2}{}^a\, \rho^{b_1b_2 c} = 0\,,\qquad 
 4\,f_{d_1d_2}{}^{[a_1}\,\rho^{a_2]d_1[b_1}\,\rho^{b_2]cd_2}
 + 3\,f_{d_1d_2}{}^{[b_1}\,\rho^{b_2c]d_1}\,\rho^{a_1a_2d_2} = 0 \,.
\label{eq:gCYBE}
\end{align}
The second requirement is a natural generalization of the homogeneous CYBE while the first one is intrinsic to the $\cE_n$ algebra [which corresponds to \eqref{alg-F}]. 
Again, the $\Omega$-field after the deformation is given by 
\begin{align}
 \Omega'^{i_1i_2i_3} = \Omega^{i_1i_2i_3} + e_{a_1}^{i_1}\,e_{a_2}^{i_2}\,e_{a_3}^{i_3}\,\bigl(\Pi^{a_1a_2a_3}+\rho^{a_1a_2a_3}\bigr)
 = \Omega^{i_1i_2i_3} + \rho^{a_1a_2a_3}\,v_{a_1}^{i_1}\,v_{a_2}^{i_2}\,v_{a_3}^{i_3}\,.
\end{align}
Namely, once we have found a solution of the generalized CYBE \eqref{eq:gCYBE}, we can easily obtain the deformed background without computing the matrix $M_A{}^B$\,. 
If we could show that the non-Abelian $U$-duality transformation \eqref{eq:NAUD} is a solution generating transformation in EFT, this tri-Killing deformation is also a solution transformation in EFT. 
In order to consider concrete applications, it is important to classify the solutions of the generalized CYBE \eqref{eq:gCYBE}. 

The tri-Killing deformation can be understood as the M-theory uplift of the YB deformation in type IIA theory, where the parameter $\rho^{a_1a_2a_3}$ is related to the $r$-matrix as $\rho^{\dot{a}\dot{b}z}=r^{\dot{a}\dot{b}}$\,. 
Then, the first equation in \eqref{eq:gCYBE} is reduced to the unimodularity condition $f_{\dot{b}_1\dot{b}_2}{}^{\dot{a}}\, r^{\dot{b}_1\dot{b}_2} = 0$\,.\footnote{The Leibniz identity \eqref{eq:unimodular} for the dual structure constant \eqref{eq:f-tilde-YB} also reproduces the same condition.}
This unimodularity condition is precisely the condition for the YB-deformed background to satisfy the supergravity equations of motion \cite{Borsato:2016ose}. 
However, the Lie algebra of the Drinfel'd double itself is consistently defined even when the unimodularity is violated. 
Moreover, as it is shown in \cite{Borsato:2016ose}, even in the non-unimodular case, the YB-deformed background does satisfy the equations of motion of the generalized supergravity \cite{1511.05795,1605.04884}, and it is also a solution of DFT \cite{1611.05856}. 
Then, a natural question is why the non-unimodular cases are excluded from the tri-vector deformation based on the $\cE_n$ algebra. 
This can be understood as follows. 

In the case of non-unimodular YB deformations, the deformed geometries are solutions of the generalized supergravity, which means that the dilaton in type IIA theory acquires a dependence on the dual coordinates $\tilde{x}_m = y_{mz}$ \cite{1703.09213}. 
In the case of DFT, the dilaton is not contained in the generalized frame fields $E_A{}^M$ and this does not cause any problem in realizing the algebra of the Drinfel'd double as $\gLie_{E_A}E_B{}^M=-\cF_{AB}{}^C\,E_C{}^M$\,. 
However, in EFT, the dilaton is contained in the generalized frame fields $E_A{}^I$ and the dual-coordinate dependence conflicts with our assumption \eqref{eq:M-section}. 
This will be the reason why the Drinfel'd double associated with non-unimodular YB deformation cannot be embedded into the $\cE_n$ algebra. 
In order to study the non-unimodular YB deformation in the context of EFT, it may be necessary to deform the $\cE_n$ algebra by changing the choice of the section \eqref{eq:M-section}.\label{page:deformed-En} 
Such deformation of the $\cE_n$ algebra may be realized also by considering a deformation of the generalized Lie derivative as it has been considered in \cite{1604.08602,1708.02589}, because the introduction of the dual-coordinate dependence is equivalent to the introduction of the deformation parameters. 

\paragraph{Connection with mathematics:}
It is interesting to investigate connections with various known facts in the mathematical literature. 
As we have mentioned, the $\cE_n$ algebra is related to the Nambu--Poisson tensor, and various results on the Nambu--Poisson group (see for example \cite{math:9812064,math:9901047,math:0204310}) will be useful to clarify the structure of the $\cE_n$ algebra. 
In addition, the $\cE_n$ algebra is a Leibniz algebra (rather than a Lie algebra), and it seems to have an intricate global structure. 
The detailed study of such global structure is also an interesting future direction. 

\subsection*{Note added}
Soon after this manuscript appeared on arXiv, the paper \cite{1911.07833} appeared, which proposes the so-called exceptional Drinfel'd algebras by using the $\SL(5)$ EFT. 
Although their algebra has not been written down explicitly in a similar form as \eqref{eq:En-algebra}, by comparing their generalized frame fields with our \eqref{eq:gen-frame} we can see that they are considering a more general ansatz,
\begin{align}
 (E_A{}^I) = \begin{pmatrix} e_a^i & 0 \\ -\tfrac{\Pi^{a_1a_2 b}\,e_b^i}{\sqrt{2!}} & \alpha\,r^{[a_1}_{[i_1}\,r^{a_2]}_{i_2]} \end{pmatrix} ,
\end{align}
where $\alpha=\alpha(x^i)$ is a certain function that has been chosen as $\alpha=1$ in this paper.\footnote{We would like to thank Emanuel Malek for useful discussion clarifying the relation between the two papers.} 
If we suppose that the $\alpha$ has a linear dependence on $x^i$\,, our $\cE_n$ algebra \eqref{eq:En-algebra} will be deformed as
\begin{align}
\begin{split}
 T_a\circ T_b &= f_{ab}{}^c\,T_c \,,
\\
 T_a\circ T^{b_1b_2} &= f_a{}^{b_1b_2c}\,T_c + 2\,f_{ac}{}^{[b_1}\, T^{b_2]c} - Z_a\,T^{b_1b_2} \,,
\\
 T^{a_1a_2}\circ T_b &= -f_b{}^{a_1a_2 c}\,T_c + 3\,f_{[c_1c_2}{}^{[a_1}\,\delta^{a_2]}_{b]} \,T^{c_1c_2} + 3\,Z_{[b}^{\vphantom{a_1}}\,\delta^{a_1a_2}_{c_1c_2]}\,T^{c_1c_2}\,,
\\
 T^{a_1a_2}\circ T^{b_1b_2} &= -2\, f_d{}^{a_1a_2[b_1}\, T^{b_2]d}\,,
\end{split}
\end{align}
where $Z_a$ are constants that correspond to $D_a\ln\alpha$. 
This deformed $\cE_n$ algebra appears to correspond to the exceptional Drinfel'd algebras of \cite{1911.07833} ($Z_a$ corresponds to their $\frac{\tau_{a5}-I_a}{3}$). 

More generally, we can make an additional twist $U_A{}^B(x^I)$ to our matrix $E_I{}^A$ as
\begin{align}
 E'_I{}^A \equiv E_I{}^B\,U_B{}^A\,.
\end{align}
Then, we find that the generalized flux, defined as $\gLie_{E'_A} E'_B{}^I = X'_{AB}{}^C\,E'_C{}^I$, becomes
\begin{align}
 X'_{AB}{}^C &= (U^{-1})_A{}^E\,(U^{-1})_B{}^F\,U_G{}^C\,X_{EF}{}^G
\nn\\
 &\quad + 2\,(U^{-1})_{[A}{}^D\,D'_{B]} U_D{}^C - Y_{FB}^{CE}\,(U^{-1})_A{}^D\,D'_E U_D{}^F\,,
\end{align}
where $D'_A \equiv E'_A{}^I\,\partial_I$\,. 
By the construction of our twist matrix, $U_A{}^B=\delta_A^B$ should be satisfied at the identity $g=1$\,. 
Then, we obtain a deformed $\cE_n$ algebra
\begin{align}
 \cF'_{AB}{}^C = \cF_{AB}{}^C + 2\, Z_{[AB]}{}^C - Y_{EB}^{CD}\, Z_{AD}{}^E \,,
\end{align}
where $Z_{AB}{}^C\equiv Z_A^{\hat{\alpha}}\,(t_{\hat{\alpha}})_B{}^C$ are certain constants that correspond to $(U^{-1})_A{}^D\,D'_B U_D{}^C$\,. 
This kind of deformation is precisely the deformation of the $\cE_n$ algebra discussed on page \pageref{page:deformed-En}, and will be useful to discuss non-unimodular YB deformations. 
Specifically, a choice $U(x^I)=\Exp{c^i \tilde{K}^z{}_z\,y_{iz}}$ ($c^i$ constant) will be required for the discussion of non-unimodular YB deformations. 

\subsection*{Acknowledgments}

This work is supported by JSPS Grant-in-Aids for Scientific Research (C) 18K13540 and (B) 18H01214.

\end{document}